\documentclass{article}

\usepackage{arxiv}
\usepackage[dvipsnames]{xcolor}
\usepackage[utf8]{inputenc}
\usepackage[T1]{fontenc}
\usepackage{hyperref}
\usepackage{url}
\usepackage{booktabs}
\usepackage{lscape}
\usepackage[raggedrightboxes]{ragged2e}
\usepackage{multirow}
\usepackage{threeparttable}
\usepackage{amsfonts}
\usepackage{amsmath}
\usepackage{mathtools}
\usepackage{nicefrac}
\usepackage{microtype}
\usepackage{lipsum}
\usepackage{graphicx}
\graphicspath{{graphics}}
\usepackage{csquotes,xpatch}
\usepackage[
        backend=biber,
        style=authoryear-comp,
        uniquename=false,
        sorting=ynt,
    ]{biblatex}

\usepackage[automake, nomain, acronym, xindy, shortcuts=ac]{glossaries-extra}
\makeglossaries
\setabbreviationstyle[acronym]{long-short}
\glssetcategoryattribute{acronym}{nohyperfirst}{true}
\newacronym{aru}{ARU}{automatic recording unit}
\newacronym{ssl}{SSL}{self-supervised learning}
\newacronym{bilstm}{BiLSTM}{bidirectional long short-term memory network}
\newacronym{ast}{AST}{audio spectrogram transformer}
\newacronym{vit}{ViT}{vision transformer}
\newacronym{auc}{AUC}{area under the curve}
\newacronym{auc roc}{AUC ROC}{area under the receiver operating characteristic curve}
\newacronym{roc}{ROC}{receiver operating characteristic}
\newacronym{stft}{STFT}{short time Fourier transform}
\newacronym{fft}{FFT}{fast Fourier transform}
\newacronym{nlp}{NLP}{natural language processing}
\newacronym{cnn}{CNN}{convolutional neural network}
\newacronym{ldc}{LDC}{Linguistic Data Consortium}
\newacronym{mfcc}{MFCC}{mel frequency cepstral coefficient}
\newacronym{knn}{k-NN}{k-nearest neighbours}
\newacronym{elev}{EV}{Elephant Voices}
\newacronym{elp}{ELP}{Elephant Listening Project}
\newacronym{dt}{DT}{decision tree}
\newacronym{rf}{RF}{random forest}
\newacronym{lr}{LR}{logistic regression}
\newacronym{mlp}{MLP}{multi-layer perceptron}
\newacronym{svm}{SVM}{support vector machine}
\newacronym{xgb}{XGB}{XGBoost}
\newacronym{dft}{DFT}{discrete Fourier transform}
\newacronym{lda}{LDA}{linear discriminant analysis}
\newacronym{lpc}{LPC}{linear predictive coding}
\newacronym{hmm}{HMM}{hidden Markov model}
\newacronym{pam}{PAM}{passive acoustic monitoring}
\newacronym{rnn}{RNN}{recurrent neural network}
\newacronym{map}{mAP}{mean average precision}
\newacronym{ap}{AP}{average precision}
\newacronym{pca}{PCA}{principal component analysis}
\newacronym{resnet}{ResNet}{residual neural network}
\newacronym{iou}{IoU}{intersection over union}
\newacronym{beats}{BEATs}{bidirectional encoder representation from audio transformers}
\newacronym{iucn}{IUCN}{International Union for Conservation of Nature}
\newacronym{rbf}{RBF}{radial basis function}
\newacronym{umap}{UMAP}{uniform manifold approximation and projection}
\newacronym{gpu}{GPU}{graphics processing unit}

\newglossaryentry{ast-seq}{
  name={AST-seq},
  description={},
}

\newglossaryentry{ast-lab}{
  name={AST-cls},
  description={},
}

\newglossaryentry{lr-frame}{
  name={LR-frame},
  description={},
}

\usepackage{float}
\usepackage{caption}
\usepackage{subcaption}
\usepackage[xindy]{imakeidx}
\makeindex

\usepackage[capitalise, noabbrev, nameinlink]{cleveref}

\usepackage[most]{tcolorbox}
\newtcolorbox{notebox}{
  parbox=false,
  colback=gray!15!white,
  colframe=gray!75!black,
  fonttitle={\scriptsize\scshape},
  enhanced,
  attach boxed title to bottom right={yshift=3mm,xshift=-4mm},
  title={Note},
}

\usepackage{doi}

\usepackage{siunitx}
\DeclareSIUnit{\unitless}{\relax}

\title{Learning to rumble: Automated elephant call classification, detection and endpointing using deep architectures}

\author{ 
\href{https://orcid.org/0000-0003-0691-0235}{\includegraphics[scale=0.06]{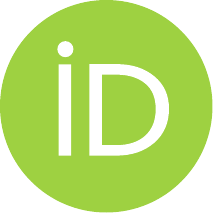}\hspace{1mm}Christiaan M. Geldenhuys}\\
	Department of Electrical and Electronic Engineering\\
	University of Stellenbosch\\
	\href{mailto:cmgeldenhuys@sun.ac.za}{\texttt{cmgeldenhuys@sun.ac.za}} \\
	\And
	\href{https://orcid.org/0000-0002-7341-1017}{\includegraphics[scale=0.06]{orcid.pdf}\hspace{1mm}Thomas R. Niesler} \\
	Department of Electrical and Electronic Engineering\\
	University of Stellenbosch\\
	\href{mailto:trn@sun.ac.za}{\texttt{trn@sun.ac.za}} \\
}

\renewcommand{\shorttitle}{\textsc{\scriptsize Automated elephant call classification and event detection using deep architectures}}

\hypersetup{
pdftitle={},
pdfauthor={Christiaan M.~Geldenhuys, Thomas R.~Niesler},
pdfkeywords={},
}

\begin{document}
\maketitle

\begin{abstract}
We consider the problem of detecting, isolating and classifying elephant calls in continuously recorded audio.
Such automatic call characterisation can assist conservation efforts and inform environmental management strategies.
In contrast to previous work in which call detection was performed at a segment level, we perform call detection at a frame level which implicitly also allows call endpointing, the isolation of a call in a longer recording.
For experimentation, we employ two annotated datasets, one containing Asian and the other African elephant vocalisations.
We evaluate several shallow and deep classifier models, and show that the current best performance can be improved by using an audio spectrogram transformer (AST), a  neural architecture which has not been used for this purpose before, and which we have configured in a novel sequence-to-sequence manner.
We also show that using transfer learning by pre-training leads to further improvements both in terms of computational complexity and performance.
Finally, we consider sub-call classification using an accepted taxonomy of call types, a task which has not previously been considered.
We show that also in this case the transformer architectures provide the best performance.
Our best classifiers achieve an average precision (AP) of 0.962 for framewise binary call classification, and an area under the receiver operating characteristic (AUC) of 0.957 and 0.979 for call classification with 5 classes and sub-call classification with 7 classes respectively.
All of these represent either new benchmarks (sub-call classifications) or improvements on previously best systems.
We conclude that a fully-automated elephant call detection and subcall classification system is within reach.
Such a system would provide valuable information on the behaviour and state of elephant herds for the purposes of conservation and management.
\end{abstract}

\section{Introduction}
\label{sec:intro}
The African bush elephant~(\textit{Loxodonta africana}) and Forest elephant~(\emph{L. cyclotis}) have been identified as endangered and critically endangered species, respectively~\autocite{iucn2020loxodontaafricana,iucn2020loxodontacyclotis}.
The rapid decline in their population is primarily due to habitat loss and to poaching.
The African bush elephant is found predominantly in the open savanna, grassland and shrubland of sub-Saharan Africa, where they rely on vast territories to roam and forage for food.
In contrast, the forest elephant inhabit the rainforests of Central and West Africa.

The Asian elephant~(\textit{Elephas maximus}) has also been identified as endangered by the \ac{iucn}.
While this species is also hunted for its ivory, it faces the particular challenge of human-elephant conflict~\autocite{iucn2020loxodontacyclotis}.
The Asian elephant is found primarily in the dry forest and grassland of Southeast Asia, where it shares its habitat with rapidly growing human settlement.
As these settlements expand, natural habitat is increasingly fragmented and reduced.
This has lead to increased competition for resources, with ensuing conflict between human and elephant.

The detection and classification of elephant vocalisations can provide crucial insights into the behaviour, distribution, and conservation status of these animals.
Automated behavioural classification can support passive monitoring and ecological management of wildlife in reserves and sanctuaries.
This high-level information can be useful for bioacoustic and ecology research a like; as well as acting as a real-time early warning system for illegal poaching activities.
Automated classification is required by further downstream task such as passive sound source localisation~\autocite{geldenhuys2023meng}.

\subsection{Contributions}
In this study, we propose an automated system for the detection and classification of elephant calls using deep learning architectures.
We include a transformer-encoder architecture, which utilises a learnable embedding token to distinguish between different call types, and has to our knowledge, not been used to process elephant vocalisations before.
To develop our models, we make use of two corpora containing labelled recordings of both African and Asian elephant rumbles.
We believe that this is the first work to use both out-of-domain and in-domain transfer learning to detect and classify elephant calls, and the first to perform explicit call segmentation.
Finally, we also perform explicit subcall classification, as a first step towards automated elephant behavioural classification.

\section{Elephant vocalisations}
African elephants have a complex vocal repertoire, consisting of various call types that serve different communicative functions and appear in different behavioural contexts.
These calls can be divided into two main categories: laryngeal calls, which originate in the larynx, and trunk calls, which are produced by a blast of air through the trunk~\autocite{poole1988social,langbauer2000elecomms,poole2005vocal,poole2011subcall}.
These call types can be further distinguished based on their acoustic measurements, such as duration, minimum frequency, and spectral characteristics~\autocite{poole2005vocal,poole2011subcall}.
By examining how elephants use these calls in different behavioural contexts, we can gain insights into the complex dynamics of their social organisation and communication.

Laryngeal calls include rumbles, roars, grunts, cries, and idiosyncratic calls such as croaks and truck-like calls. Rumbles are low-frequency vocalisations that can travel long distances and are used in a variety of contexts, including greeting, reuniting, and reassurance.
They can also convey information about the caller's identity, age, and reproductive status~\autocite{soltis2010vocal}.
Roars are loud, high-frequency calls that serve to intimidate or warn other animals, while grunts are soft, low-frequency vocalisations used in close-range communication.
Cries are high-pitched, urgent calls produced in response to separation or danger~\autocite{poole1988social}.
The husky cry is a variant of the cry with a rougher, more strained quality.
Croaks and truck-like calls are structurally unique calls that may not be socially relevant, with limited exemplars having been observed~\autocite{poole2011subcall}.

Trunk calls include trumpets, bursts, snorts, and chirps.
Trumpets are loud, high-frequency vocalisations produced by exhaling through the trunk, often used in greeting, excitement, or as a contact call~\autocite{poolegranli2011behaviour}.
Bursts are short, explosive calls produced by rapidly expelling air through the trunk, while snorts are similar but less intense.
Chirps are high-frequency, bird-like calls produced by vibrating the tip of the trunk~\autocite{soltis2010vocal}.
The nasal trumpet is a variation of the trumpet call produced through the nose instead of the mouth, resulting in a softer, more muffled sound~\autocite{poole2011subcall}.

\section{Classification models}
\label{sec:background}
We will compare the performance of various classifier architectures when applied to the task of elephant call classification, detection and endpointing.
In the following sections we introduce these techniques.

\subsection{Shallow classification models}
\label{sec:background:shallow}
In addition to current neural network classifiers, three shallow architectures were considered as baselines.
These are described briefly in the following: \acf{lr}, \acfp{svm} and gradient boosting (specifically \acl{xgb}).

\subsubsection{Logistic regression}
\Ac{lr} is a well-established linear approach to binary classification problems.
The model estimates the log odds of an event's occurrence as a linear combination of the input features.
The sigmoid function is then applied to the log odds in order to obtain an probabilistic output, used for classification.
The posterior class probabilities are optimised through maximum likelihood estimation of the model parameters.
\Ac{lr} is popular due to its simplicity, interpretability, and strong performance when the independent variables are linearly related to the log odds of the dependent variable.

\subsubsection{Support vector machines}
\Acp{svm} aim to find a hyperplane decision boundary that best separates data points of different classes~\autocite{cortes1995svm}.
This hyperplane is typically chosen as the one with the largest margin, which is the distance between the hyperplane and the closest data points from each class, also known as the support vectors.
\Acp{svm} use different kernel function transformations to transform data from a space in which they might not be linearly separable into a different space in which they are.
Commonly used kernel functions include linear, polynomial, radial basis function, and sigmoid kernels.

\subsubsection{Gradient boosting}
Decision trees recursively partition data into subsets based on feature values, leading to a tree-like structure.
At each internal node of the tree, a test on an attribute is performed, and branches lead away based on the outcome of this binary test (i.e. true or false).
The leaf nodes of the tree represent class labels in classification tasks.
Decision trees can suffer from overfitting due to their tendency to create complex decision structures that memorise training samples rather than generalise patterns within the dataset \autocite{quinlan1986decisiontrees,breiman2001randomforest}.
To address this, various pruning strategies and tree ensemble methods have been proposed, such as reduced error pruning, cost complexity pruning, and random forests.

Random forests~\autocite{breiman2001randomforest} are an ensemble learning method.
A random forest consists of multiple decision trees trained on different sub-samples of the original dataset.
Each of the trees uses a random subset of features to split nodes.
The final prediction is made by averaging the predictions of all individual trees in the forest.

Gradient Boosting represent a popular ensemble technique that combines multiple weak performing models~(e.g.\ decision trees) to build an overall model that preforms better than the individual weak component models.
The primary goal is to improve the ensemble model estimation through sequential learning based on the weaknesses of previous models.
XGBoost~\autocite{chen2016xgb} uses a tree-based algorithm that focuses on minimising the objective function by iteratively fitting trees to the residual errors from the previous step.

\subsection{Deep neural classification models}
\label{sec:background:deep}
The shallow classifiers described above are well-established and have already seen application in the field of elephant call detection~\autocite{clemins2003elepspeech,zeppelzauer2015autoclass,silva2017waveletthesis}.
Recently, neural architectures have come to represent the state-of-the-art in many machine learning applications.
The following section describes three deep learning architectures, namely \acfp{mlp}, \acfp{cnn} and transformer-based~(encoder-only) architectures.
Furthermore, it describes the concepts of transfer learning and self-supervised learning through self-distillation.

\subsubsection{Multi-layer perceptron}
\Acp{mlp}~\autocite{rumelhart1986mlp} are feedforward neural networks, consisting of multiple layers of neurons that process input data through a series of hidden layers to produce a final model estimate.
Each neuron is fully-connected to all neurons of the preceding layer and includes a non-linear activation function.
Similar to \ac{lr} models, \acp{mlp} are trained by maximising the posterior class probability.

\subsubsection{Convolutional neural network}
\Acp{cnn}~\autocite{fukushima1980cnn,lechun1989cnn} are a class of deep learning models traditionally applied in the field of computer vision and image processing.
Originally inspired by the biological structure of the visual cortex, \acp{cnn} were designed to mimic human vision.
The a \ac{cnn} architecture includes convolutional layers, pooling layers, and fully connected layers.
Convolutional layers use small kernel filters to extract local features from the input, pooling layers downsample the feature maps to reduce dimensionality and computational cost, and fully connected layers enable the network to model complex decision boundaries.
Over the last two decades, significant advances have been made in \ac{cnn} architecture, such as AlexNet~\autocite{krizhevsky2012alexnet}, VGGnet~\autocite{liu2015vggnet}, and \acp{resnet}~\autocite{he2016resnet}.
The latter employs skip connections between layers that allow deeper models to be more effectively trained.

\subsubsection{Transfer learning}
Transfer learning aims to improve model performance on a target domain by \textit{transferring knowledge} contained in a different source domain~(out-of-domain pre-training) or similar source domain~(in-domain pre-training)~\autocite{pan2009transfer}.
Transfer learning has proven to be especially useful in low-resource or data sparse settings~\autocite{zhuang2020transfer}, as deep neural architectures typically require large quantities of data.

Following pre-training, the model can be \textit{fine-tuned} on the domain-specific data.

\subsubsection{Transformers}
The transformer architecture was first proposed by \autocite{ashish2017attention} as a successor to \ac{rnn} for sequence-to-sequence \ac{nlp} tasks.
Transformers have an encoder-decoder structure that relies on the attention mechanism to model both long and short term sequence dependencies.
The architecture benefits from reduced training times, due to the consistent number of operations, compared to the recurrent networks perform a variable number of operations, dependant on the input sequence length.

As a result, the transformer architecture can be parallelised during training, while \acp{rnn} have to be trained sequentially.
Furthermore, the global attention mechanism allow transformers to benefit from a constant dependence on path-length, which allows improved long term and short term dependency modeling.
However, this comes at the cost of a memory requirement that is quadratic with sequence length.

\subsubsection{Vision transformers}
The standard transformer architecture, as proposed by \textcite{ashish2017attention}, was intended for machine translation and has since become the dominant method in \ac{nlp}.
A na\"{i}ve application of the self-attention mechanism to images would require that each pixel attends to every other pixel.
However, the quadratic memory requirement of the global attention mechanism, in terms of input sequence length, would make this practically infeasible for images.
Thus, to apply transformers~(encoder only) in the context of image processing, several modifications to the global attention mechanism have been suggested~\autocite{parmar2019vitalt1,hu2018vitalt2,zhoa2020vitalt3}.
\Textcite{dosovitskiy2020vit} propose that a single two-dimensional image is divided into a sequence of smaller images, known as \textit{patches}.

A trainable linear projection is used to reduce the dimensionality of each patch into a fixed size latent vector, known as a \textit{patch embedding}.

The process of obtaining a patch embedding is similar to the textual embedding process commonly employed in traditional \ac{nlp} tasks.
Similar to standard transformers, a positional embedding or encoding scheme~\autocite{ashish2017attention} is applied to each patch embedding.
This positional encoding remains one-dimensional, as two-dimensional encoding schemes have not proven more effective~\autocite{dosovitskiy2020vit}.
An optional learnable embedding token \texttt{[CLS]}, can be appended to the start of the patch embedding sequence, the output of which serves as the model classification estimate~\autocite{devlin2019bert}.
The \ac{vit} architecture allows image classification to make effective use of the self-attention mechanism, employed by the transformer-encoder architecture.

\subsubsection{Audio spectrogram transformer}
\label{sec:background:ast}
\Textcite{gong2021ast} proposes the \ac{ast} architecture, which performs audio classification by applying a \ac{vit} to a two-dimensional time-frequency spectral representation~(mel-spectrogram) of an input audio sequence.
Applying deep neural vision classification models to audio classification was first proposed by \textcite{hershey2017cnnaudio}.
Here the authors applied a \ac{cnn} architecture typically used for image classification to a two-dimensional representation of an input audio sequence.
The \ac{ast} architecture first converts the input audio waveform into a 128-dimensional log-mel spectrogram representation, computed using a \qty{25}{\milli\second} Hamming window with a stride of \qty{10}{\milli\second}~(a typical configuration for speech recognition tasks).
The spectrogram is then split into $16 \times 16$ patches and passed through a \ac{vit} model, with an embedding size of 768~dimensions, 12~transformer encoding layers each with 12~self-attention heads~\autocite{gong2021ast}.
As in the \ac{vit} architecture, a learnable embedding token \texttt{[CLS]}, is prepended the patch embedding sequence, the output of which is used to perform sequence classification.

\subsubsection{Self-supervised pre-training}
Annotating and labelling datasets is expensive because it typically requires human expertise.

This makes techniques that do not depend on target labels attractive.
\textit{Self-supervised learning} is an approach where the target labels are either generated by the source data or using another algorithmic system.
Self-supervised learning allows supervised training methods to be used on data previously only accessible to unsupervised techniques.
Commonly, models are pre-trained using self-supervision, in order to obtain semantic knowledge of the underlying distribution, that may aid in further downstream tasks.
These pre-trained models are then typically fine-tuned on a supervised dataset for which target labels are available.
This fine-tuning dataset can be orders of magnitude smaller than the datasets used for pre-training.
This allows models that were previously intractable to be employed.
However, additional care should be taken to ensure the model does not overfit to the underlying distribution of data.

\subsubsection{Bidirectional encoder representation from audio transformers}
\label{sec:background:beats}
\Textcite{chen2022beats} presents an \ac{ssl} technique for iterative audio pre-training of bidirectional encoder architectures, such as transformer-encoders and \acp{bilstm}.
The method consists of a teacher-student configuration in which semantic knowledge about the data is obtained, through iterative knowledge distillation, without the use of target labels.
The method starts with a tokeniser in the form of a set of codebook embeddings that have been randomly initialised and therefore contain no learned information.
The input spectrogram is divided into $16 \times 16$ patches and passed through a learnable projection layer.
The projected patches are compared using a nearest neighbour match with the codebook embeddings.
The retrieved codebook embedding is used as the target label for the bidirectional encoder (teacher model).

The bidriectional encoder is then trained to map the given input spectrogram to the target labels, produced by the tokenizer.
As for the tokeniser, the input spectrogram is divided into patches and passed through a linear projection layer.
A random attention masking scheme is applied to the resulting patch embeddings, obscuring both features in time and frequency.
Only the unmasked patch embedding are passed to the bidirectional encoder to obtain a latent representation of the respective unmasked patches.
These unmasked latent representations, along with specific masked tokens in place of the masked latent representation, \texttt{[M]}, are passed to the label prediction layer, typically a single linear layer with a sigmoid activation function.
The predicted label, for each masked token, is compared with the target label generated by the tokeniser through a masked loss function, such as masked cross entropy.
Hence the loss is only evaluated between target labels and masked tokens.
This in turn forces the model to embed semantic information about the predicted unmasked tokens within the masked tokens, while only using a small subset of the input features.
Once the bidirectional encoder has been trained, a new set of target labels is generated and used as ground truth values for the tokeniser model, discarding the previous tokeniser's target labels.

During subsequent iterations, the tokeniser codebook embeddings are trained on the new output labels from the encoder model.
The new tokeniser target labels are compared to the teacher target labels through cosine similarity, and a set of learnable codebook embeddings are obtained.
Once the tokeniser has been sufficiently trained, a new set of target labels is generated, which is used to retrain the encoder model.

This process of training the encoder model using the tokeniser labels and distilling the encoder predictions into the tokenizer is repeated several times for a large unlabelled audio dataset, such as \textit{AudioSet-2M}\footnotemark~\autocite{gemmeke2017audioset}, and is referred to as self-distillation~\autocite{chen2022beats}.
\footnotetext{
  AudioSet is a large-scale dataset consisting of over 2 million 30-second sound clips, each labeled with one or more of 527 audio event classes.
  It has been widely used in audio classification and self-supervision tasks.
  The dataset covers a wide range of categories, from music and speech to nature sounds and human activities.
}
The result is a encoder model that has been pre-trained to estimate a set of quantised target labels from an unlabelled dataset.
This pre-trained model can then be fine-tuned on a domain-specific source set, through transfer learning.

\section{Literature Review}
\label{sec:lit-review}
In the following, we review the literature associated with two distinct objectives, namely \emph{elephant call classification} and \emph{elephant call detection}.
Both fall within the realm of bioacoustic classification, but have received limited attention in the research literature.
\Cref{tab:lit:results-summary} provides the quantitative summary of the results for the literature considered in this review.

\emph{Call detection} is the task of determining whether a particular audio signal contains an elephant call or not.
A related task to call detection is \emph{call endpointing}, where the start and end of the call is identified within an audio signal.
When call detection is performed per-frame, the call endpointing is implicitly performed.
To our knowledge, no work in the literature has focused on explicit call endpointing for elephant vocalisations.
\emph{Call classification} is a segment-level task in which a short segment of audio that has already been endpointed is considered to identify the particular type of call.
While call classification is always a multi-class task, call detection can be either binary or multi-class.
To our knowledge, currently no literature considers multi-class call detection.
Call classification can be viewed as a down stream task of either call detection or call endpointing.

\begin{table}
\centering
\caption{
    Summary of quantitative results for elephant call detection and classification reported in literature.
    Results are not directly comparable, as evaluations were performed on different datasets.
    \emph{5-class}: 5 class call classification, \emph{2-class}: 2 class call classification, \emph{Det}: binary call detection.
}
\label{tab:lit:results-summary}
\begin{tabular}{@{}lcccccc@{}}
\toprule
Literature                          & Task      & Accuracy      & Precision     & Recall/Sensitivity & Specificity   \\ \midrule
\Textcite{clemins2003elepspeech}    & 5-Class.  & 79.7          & --             & --                  & --             \\
\Textcite{venter2010elepspeech}     & Det.      & --             & 85.7          & 85.7               & --             \\
\Textcite{stoeger2012lda}           & 2-Class.  & 99.0          & --             & --                  & --             \\
\Textcite{zeppelzauer2015autoclass} & Det.      & --             & --             & 88.2               & 86.3          \\
\Textcite{keen2017elepcnn}          & Det.      & --             & --             & 87.2               & 91.1          \\
\Textcite{silva2017waveletthesis}   & Det.      & --             & 84.0          & 84.0               & --             \\
\Textcite{bjorck2019elepcnn}        & Det.      & 91.7          & 91.4          & 91.4               & --             \\
\Textcite{leonid2022elepcnn}        & Det.      & 96.2          & 93.2          & 88.2               & 87.2          \\ \bottomrule
\end{tabular}
\end{table}

\Textcite{leong2003calldef} worked on standardising the classification of African elephant calls, utilising measurements such as bandwidth, sound quality, fundamental frequency, infrasonic elements, and duration.
The authors identified eight distinct call types, including three rumble variants that were distinguished by their bandwidth.
A cross-correlation analysis was performed on the fundamental frequency contour of all rumble calls, the most common type, revealing five rumble categories.
However, multidimensional scaling showed minimal clustering among call types, implying either overlapping rumble types or that the fundamental frequency contour may not be the primary physical property conveying the meaning of these signals.

The first application of speech processing techniques to the classification of elephant vocalisation signals, that we are aware of, was conducted by \textcite{clemins2003elepspeech,clemins2005elephmm}.
These authors considered \ac{mfcc} features, manually adjusting the mel-scale to account for the infrasonic nature of elephant calls.
Along with \acp{mfcc}, the log frame energies of the recordings were used to classify elephant vocalisations using a \ac{hmm}.
The authors obtained a 5-class~(\texttt{croak}, \texttt{rumble}, \texttt{rev}, \texttt{snort} and \texttt{trumpet}) average call classification accuracy of \qty{79.74}{\percent} on a balanced set containing 74 vocalisations.
The classification accuracy improved to \qty{94.29}{\percent} when omitting noisy recordings, using the same dataset and labels as \textcite{leong2003calldef}.

\Textcite{wood2005cluster} went on to show that, by applying model-based cluster analysis to a set of acoustic features extracted from elephant rumbles, they could identity four distinct clusters when using frequency contour features and three clusters when using extracted acoustic parameters.
Each of these clusters corresponded to specific observed animal behaviour, which corresponds with the findings of \textcite{leong2003calldef}.

\Textcite{venter2010elepspeech} applied a subband pitch detector, originally developed for voice activity detection, to achieve elephant rumble detection by locating audio segments within which the pitch varied less than a predetermined threshold.
The algorithm consists of two steps, first the pitch is estimated for an entire audio recording and subsequently segments of low pitch variability is extracted.
Each extracted segment is considered a detected elephant rumbles.
The algorithm is only evaluated on a segment-level, with no results given on the boundary endpoint performance.
This algorithm achieved for both segment-level precision and recall a score of \qty{85.7}{\percent}.

\Textcite{stoeger2012lda} applied \ac{lda}, a \ac{svm} and a \ac{knn} classifier to distinguish between \textit{oral-} and \textit{nasal} rumbles.
Each classifier used features obtained from a \ac{lpc} smoothed spectrogram representation of the recordings, obtaining 99\
Unlike previous approaches that relied on call-specific characteristics (such as predefined formant frequencies), this approach was fully automatic and made no a priori assumptions about the call structure.

The work described above was performed in either a controlled or a captive setting, which may not be representative of actual field conditions.
\Textcite{zeppelzauer2013eledetmethod} and subsequently \textcite{zeppelzauer2015eledetsystem} were the first to apply automatic classification techniques to free-roaming elephants.

Two techniques were applied to elephant rumble detection, the first consisting of a \ac{svm} classifier using Greenwood cepstrum features with additional spectral signal preprocessing, and the second a template matching algorithm.

In the case of the \ac{svm} classifier, a spectral representation of the entire audio signal is obtained, 8 of the \ac{stft} frames are averaged before computing the Greenwood cepstrum features.
A classification in made once for each of these aggregated frames~(approximately every \qty{120}{\milli\second}).
The template matching algorithm achieved a sensitivity of \qty{78.6}{\percent} and a false discovery rate of \qty{21.4}{\percent}, and was outperformed by a \ac{svm} classifier, for which the corresponding figures were \qty{88.2}{\percent} and \qty{86.3}{\percent}.

\Textcite{keen2017elepcnn} proposed a set of handcrafted two-dimensional convolutional kernels~\autocite{leung2001horzcnnfeat} and, using an \ac{svm}, \ac{rf} and AdaBoost classification algorithms, improved on the state-of-the-art set by \textcite{zeppelzauer2015eledetsystem} for elephant rumble detection, achieving a sensitivity of \qty{87.2}{\percent} and specificity of \qty{91.0}{\percent}.

\Textcite{silva2017waveletthesis} goes on to propose the use of a \ac{svm} classifier with wavelet-based features, achieving a precision and recall of \qty{84.0}{\percent} on the task of elephant rumble detection.

\Textcite{leonid2022elepcnn} applied a \ac{cnn} model to public domain data some of which we shall also use in our experimental evaluation.
Achieving an average recall of \qty{94.4}{\percent}, a specificity of \qty{88.2}{\percent} and precision of \qty{89.2}{\percent}.

The work described so far has made use of manual field recordings.
\Ac{pam}, an alternative approach to data collection, has become a popular means to record large audio datasets autonomously~\autocite{wrege2017elepam}, and has enabled the use of more sophisticated classification models.
To our knowledge, \textcite{bjorck2019elepcnn} were the first to apply a deep neural network to automatic elephant call detection to \ac{pam} recordings.
Using \ac{mfcc} features and a CNN-LSTM classifier, the authors achieve, a precision of \qty{90.8}{\percent} and recall of \qty{96.4}{\percent}.

\section{Data}

Our experiments were conducted on two audio datasets that contain elephant vocalisations.
The first is a compilation of handheld field recordings of the African elephant, available in the public domain through the \textit{\ac{elev}} conservation project~\autocite{poole2021elev}.
The second is a collection of handheld field recordings of the Asian elephant, provided by the \textit{\ac{ldc}}~\autocite{ldc2010asianelevoc}.
Both datasets are of free-roaming elephant.
\Cref{tab:dataset:summary} provides a high-level comparison of the two datasets.

\begin{table}
\centering
\begin{threeparttable}
\caption{Summary of \acf{elev} and \acf{ldc} elephant vocalisations datasets used for experimentation.}
\label{tab:dataset:summary}
\begin{tabular}{@{}rcc@{}}
\toprule
Dataset                   & \textit{\acs{elev}}                      & \textit{\acs{ldc}}                         \\ \midrule
Authors                   & \textcite{poole2021elev}                 & \textcite{ldc2010asianelevoc}              \\
Elephant species          & \textit{Loxodonta africana}              & \textit{Elephas maximus}                   \\                    
Recording environment     & Handheld field recordings                & Handheld field recordings                  \\
Recording equipment       & ARES-BB Nagra                            & Fostex FR-2                                \\
Microphone                & Not specified                            & Earthworks QTC50                           \\
Number of call types      & 40                                       & 14                                         \\
Detail of annotation      & File-level                               & Within \qty{100}{\milli\second}            \\
Total recording duration  & \qty{1}{hour}                            & \qty{57.5}{hours}                          \\
Of which annotated        & \qty{36}{minutes}\tnote{\ddag}           & \qty{5.4}{hours}                           \\
Number of recordings      & 226                                      & 1577                                       \\
Average length            & \qty{14.75}{\second}                     & \qty{131.03}{\second}                      \\
Min.                      & \qty{0.49}{\second}                      & \qty{1.25}{\second}                        \\
Max.                      & \qty{296.52}{\second}                    & \qty{3889.38}{\second}                     \\
Std. dev.                 & \qty{31.48}{\second}                     & \qty{177.02}{\second}                      \\
Number of vocalisations   & 514                                      & 4433                                       \\
Sampling rate             & \qty{44.1}{\kilo\hertz}                  & \qty{16}{\kilo\hertz}                      \\
Bit depth                 & 16-bit                                   & 24-bit                                     \\
Number of channels        & 2                                        & 1\tnote{\dag}                              \\
Low frequency cut-off     & Not specified                            & \qty{3}{\hertz}                            \\ \bottomrule
\end{tabular}
\begin{tablenotes}
    \scriptsize
    \item[\dag] The second channel is used for field notes.
    \item[\ddag] Annotations provided by authors, as described in \cref{sec:dataset:elev}.
\end{tablenotes}
\end{threeparttable}
\end{table}

\subsection{Elephant voices project (EV dataset)}
\label{sec:dataset:elev}
Over the course of several years, \textcite{poole2021elev} have collected a set of recordings of African elephant~(\textit{Loxodonta cyclotis}) vocalisations.
The recordings were made in three locations: Amboseli and Maasai Mara in Kenya, and Gorongosa National Park in Mozambique.
A subset of these recordings (approx. 230) is available in the public domain.
This subset has been annotated according to an ethogram presented in \textcite{poole1988social,poole1994sex}.
Each recording is accompanied by a single elephant call type annotation indicating the dominant or overarching call type.
However, multiple different calls may be present in a recording.
This data can therefore be regarded as \textit{weakly} labelled, since there is only a single label per recording and there is no temporal information indicating the exact time of occurrence of the call.
To address this, we have added temporal labels to this dataset by manually identifying the start and end of each elephant vocalisation.
The addition of these \textit{strong} labels allows the dataset to be used not only for elephant call classification, but also for call detection and endpointing.
The resulting time-labelled dataset contains 922~unique vocalisations spanning \qty{36}{minutes} of audio in 230~recordings of the total 1-hour recording duration.
All recordings were made using a ARES-BB Nagra recorder.
However, the model of the microphone used is not specified.
This dataset is similar to the data used by \textcite{leonid2022elepcnn}, thus making results the reported comparable to the findings in their study.

\subsection{Asian elephant vocalisations (LDC dataset)}
\label{sec:dataset:ldc}
The \acf{ldc} hosts a dataset of Asian elephant~(\textit{Elephas maximus}) vocalisations recorded and annotated by \textcite{ldc2010asianelevoc}.
The dataset consist of \qty{57.5}{hours} of Asian elephant vocalisation recordings, made in the Uda Walawe National Park, Sri Lanka; of which \qty{31.25}{hours} have been annotated.
The vocalisations are identified based on those that show clear fundamental frequencies~(periodic), those that do not~(aperiodic) and those that show both periodic and aperiodic regions.
In this way, 14~unique call types can be identified~\autocite{mckay1973asianelephant}.
The dataset is strongly labelled and includes the diarisation for up to four overlapping elephant vocalisations.
To allow accurate diarisation, the dataset is annotated in a \textit{multi-label} fashion, which allows more than one call type to occur at the same time.
Field recordings were made using a Fostex FR-2 field recorders and an Earthworks QTC50 microphone~(capable of recording infrasonic signals down to \qty{3}{\hertz}).

\subsection{Target label annotations} \label{sec:target-label}
The \ac{elev} and \ac{ldc} datasets have been annotated using different labelling schemes~\autocite{poole1988social,mckay1973asianelephant}.
These differences do not only relate to the terminology used by the authors, but also to intrinsic differences in the ecological behaviour and acoustic properties of the vocalisation of the two different elephant species.
\Textcite{langbauer2000elecomms} presents a unified annotation scheme for elephant vocalisation and compares the differences between annotation styles used in the two datasets we consider.

\Cref{tab:call-summary} summarises the different call types in each of the two datasets according to this unified annotation scheme, and specifies total call duration and number of occurrences, for each dataset respectively.

Each dataset does not contain the same range of call types.
However, both corpora contain \texttt{rumble}, \texttt{trumpet} and \texttt{roar}, allowing for comparison of classifier performance across datasets for these call types.
While both corpora contain \texttt{croak} and \texttt{squelch}, there are insufficient samples of these call types to reliably evaluate classifier performance
Furthermore, the \texttt{snort}, \texttt{nasal-trumpet}, \texttt{husky-cry} and \texttt{truck-like} call types have been omitted, due to an insufficient number of occurrences to allow cross-validation.

\begin{figure}
  \centering
  \includegraphics{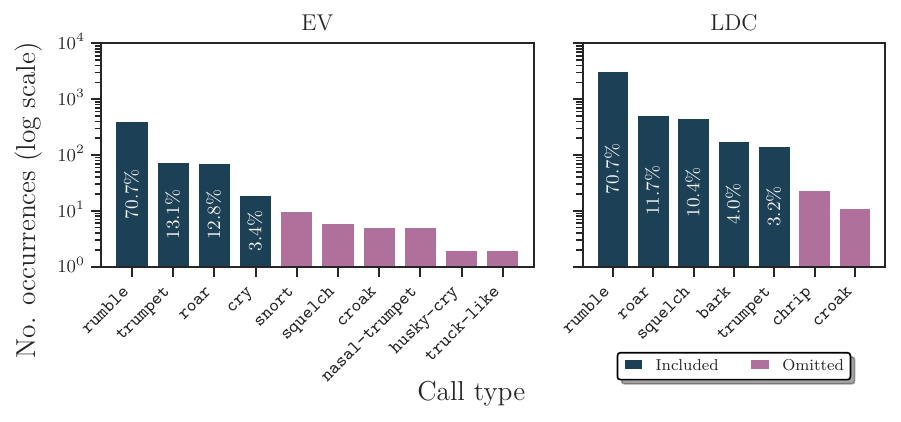}
  \caption{
    Number of occurrences of each call type in the \acf{elev} and the \acf{ldc} corpora.
    Some call types are omitted from experiments due to  an insufficient number of occurrences to allow cross-validation.
    Each included call type has been annotated with the class prevalence of each call type, within their respective dataset.
  }
  \label{fig:dataset:call_nseg}
\end{figure}

\Cref{fig:dataset:call_nseg} shows the prevalence for each respective call type in each corpus.
From this figure it clear that both datasets are severely unbalanced, with the \texttt{rumble} call type dominant in both, followed by \texttt{roar} and \texttt{squelch} in the \acs{ldc} dataset and \texttt{trumpet} and \texttt{roar} in the \acs{elev} dataset.
The \texttt{rumble} and \texttt{roar} call type occurs with approximately the same frequency in both datasets.
To address this class imbalance, performance metrics will be computed for each class independently and the average over all classes taken.
There is also a considerable difference in the number of segments, the \ac{ldc} dataset containing at least an order of magnitude more segments for each call type.

\subsection{Composite calls and Subcalls}
\label{sec:subcalls}
Elephants also produce a series of amalgamated calls, referred to as \emph{composite} calls~\autocite{poole2011subcall}.
These calls consist of a combination of two or more fundamental call types~(e.g., \texttt{rumble}, \texttt{roar}) produced as a single vocalisation~(e.g., \texttt{roar-rumble}).
Composite calls have received limited attention in literature, but have been observed to indicate excitement or disturbance~\autocite{poole2011subcall}.
For the purpose of our experiments, these calls have been annotated based on their fundamental call types, for example the start of a \texttt{roar-rumble} would be annotated as \texttt{roar} while the end would be annotated as \texttt{rumble} without any separating silence.
As the vocalisation transition seamless between the different call types, the exact transition boundary may be difficult to annotate exactly.

\begin{figure}
  \centering
  \includegraphics{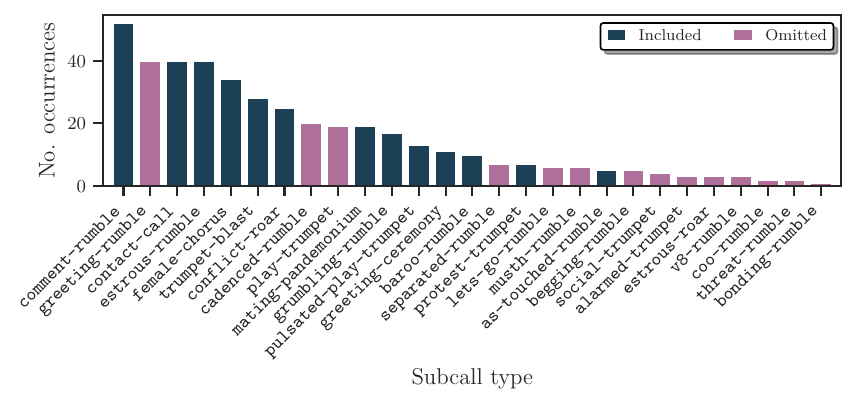}
  \caption{
    Number of occurrences of each subcall type present in the \acf{elev} corpus.
    Some subcall types are Classes omitted from experiments due to an insufficient number of occurrences to allow cross-validation.
  }
  \label{fig:dataset:subcall_nseg}
\end{figure}

Elephant do not only exhibit vocalistions that form part of their major call types, such as \texttt{rumble} and \texttt{roar}, but also have more nuanced calls referred to as \emph{subcalls}~\autocite{poole2011subcall}.
These calls accompany particular elephant behaviour during for example parent-offspring interactions, hormonal and emotional state and social interaction.
As this behaviour can not be determined based only on the major call types, automated subcall classification is the first steps to automated non-invasive elephant behavioural classification.
We perform subcall classification experiments only for the \ac{elev} dataset, as the \ac{ldc} data do not contain subcall annotations. \Cref{fig:dataset:subcall_nseg} shows the spread of the subcalls types used for experimentation.
Some subcalls are to infrequent to perform cross-validation, and have thus been omitted.

\begin{table}
\centering
\begin{threeparttable}
\caption{
Call types present in the \acf{elev} and the \acf{ldc} corpora, according to the unified annotations convention proposed by \textcite{langbauer2000elecomms}.
The number of segments for each call type, the total duration of each call type, and the associated target label are listed. 
The number of cross-validation folds is denoted by $K$ and is smaller for \acs{elev} due to its smaller size.
}
\label{tab:call-summary}
\begin{tabular}{@{}lllllll@{}}
\toprule
                                            &                   & \multicolumn{2}{c}{\textit{EV} $\left(K=5\right)$} &  & \multicolumn{2}{c}{\textit{LDC} $\left(K=10\right)$} \\ \cmidrule(lr){3-4} \cmidrule(l){6-7}
\Textcite{langbauer2000elecomms} call type  &                     & \# Segments   & Total duration [sec] &  & \# Segments   & Total duration [sec] \\
\midrule        
    \multicolumn{7}{c}{\textbf{Advertisement of hormonal state}}                                                                                   \\
\midrule        
Estrous rumble                              &                     & 40                 & 169.7           &  & --               & --                \\
Musth rumble                                &                     & 6\tnote{\dag}         & 34.2            &  & 9\tnote{$\ast$}  & 11.0              \\
\midrule                    
    \multicolumn{7}{c}{\textbf{Advertisement of emotional state}}                                                                                  \\
\midrule                    
Female chorus                               &                     & 23                 & 198.8           &  & --               & --                \\

Long roar                                   &                     & --                 & --              &  & 150              & 413.3             \\
Roar                                        &                     & 19                 & 42.1            &  & 47               & 76.0              \\
Mating pandemonium                          &                     & 18\tnote{\dag}     & 177.8           &  & --               & --                \\
Play trumpet                                &                     & 9\tnote{\dag}        & 7.5             &  & --               & --                \\
Social trumpet                              &                     & 4\tnote{$*$}      & 6.4             &  & --               & --                \\

Trumpet blast                               &                     & 27                 & 49.8            &  & --               & --                \\
Snort                                       &                     & 6\tnote{\dag}      & 7.6             &  & --               & --                \\
Trumpet                                     &                     & 28                 & 54.4            &  & 144              & 195.9             \\
Rumble                                      &                     & 162                & 695.6           &  & 183              & 1234.2            \\
Chirp-rumble                                &                     & --                 & --              &  & 23               & 77.9              \\
Croak-rumble                                &                     & --                 & --              &  & 11               & 19.8              \\
\midrule                
    \multicolumn{7}{c}{\textbf{Group cohesion and coordination}}                                                                                   \\
\midrule                
Growl\tnote{\ddag}                              &                     & --                 & --              &  & 2972             & 15476.7           \\
Let's go rumble                             &                     & 6\tnote{\dag}      & 35.5            &  & --               & --                \\
Contact rumble                              &                     & 40                 & 138.8           &  & --               & --                \\
Greeting rumble                             &                     & 40\tnote{\dag}     & 178.7           &  & --               & --                \\

\midrule            
    \multicolumn{7}{c}{\textbf{Affiliative}}                                                                                                                    \\
\midrule            
Cry                                         &                                 & 9                   & 13.17           &  & ---               & ---                \\
\midrule        
    \multicolumn{7}{c}{\textbf{Misc}}                                                                                                                           \\
\midrule        
Squelch                                     &                                 & 6                   & 11.5            &  & 43               & 53.4              \\
Squeak                                      &                                 & ---                  & ---              &  & 423              & 250.5             \\
Croaking                                    &                                 & 5\tnote{\dag}                   & 29.4            &  & ---               & ---                \\
Woosh                                       &                                & 4\tnote{$\ast$}     & 6.6             &  & ---               & ---                \\
Bark                                        &                                 & ---                  & ---              &  & 35               & 35.4              \\
Trunk call                                  &                                 & 8                   & 23.8            &  & ---               & ---                \\
Ceremony                                    &                                & 18                   & 227.4           &  & ---               & ---                \\
\midrule    
    \multicolumn{7}{c}{\textbf{Composite}}                                                                                                                       \\
\midrule    
Long roar-rumble                            &                                & ---                   & ---              &  & 254              & 1074.4            \\
Roar-rumble                                 &                                & 5                    & 15.4            &  & 75               & 255.2             \\
Bark-rumble                                 &                                & ---                   & ---              &  & 142              & 450.3             \\
Rumble-cry                                  &                                & 6                    & 10.0            &  &                  &                   \\ \midrule
Total                                       &                                & 489                  & 2134.2          &  & 4511             & 19624.1           \\ \bottomrule
\end{tabular}
\begin{tablenotes}
    \scriptsize
    \item[$\ast$] Omitted due to fewer than one segment per cross-validation fold.
    \item[\dag] Omitted due to fewer than one recording per cross-validation fold.
    \item[\ddag] Based on the findings in \textcite{silva2010asianelephant}, the Asian elephant's \textit{growl} are acoustically similar to the African elephant \textit{rumble} identified by  \textcite{poole1988social}.
\end{tablenotes}
\end{threeparttable}
\end{table}

\subsection{Cross-validation}\label{sec:kfold}
When developing machine learning-based models, separate training, testing and development datasets are required.

Since our datasets are small for most vocalisation classes, we have partitioned our data into training, testing, and development datasets with a view to cross-validation~\autocite{stone1974crossval,mosteller1968crossval}.

K-fold cross-validation partitions the data into $K$ subsets, referred to as \textit{folds}.
In a sequence of $K$ \textit{turns}, each of the folds is held out as a test set, while the remaining $K-1$ folds are used for model training.

Nested cross-validation extends this process by introducing an \emph{inner} and \emph{outer} turn.
For each outer turn, a test fold is kept aside for testing.
For each inner turn, from the remaining $K-1$ training folds a development fold is held out for parameter selection and model performance validation, while training the model on the remaining $K-2$ folds.
The best performing model, based on the $K-1$ inner-fold development sets, is chosen and retrained on the complete inner fold, while still excluding the outer (test) fold.
Hence, for each inner turn, a different fold is used as a development set.
The optimal hyperparameters for each outer turn are selected based on the average inner turn development loss.

When dividing a dataset into folds~(disjoint subsets), it is important to ensure that the class distribution remain even.
Stratification is the process of ensuring the target label distribution for a given partition of the data is representative of the overall dataset target label distribution.
The dataset is partitioned such that each fold is stratified with respect to the number of elephant call types, thereby preserving the overall call occurrence distribution within each fold.
Additionally, no single recording occurs in more than one fold.

Cross-validation is especially suitable for small datasets, because here the risk of choosing a single static test set that is biased due to sampling variation is higher than it is for large datasets.

Both datasets, in \cref{tab:dataset:summary}, are small in comparison to those typically used in audio classification~\autocite{gemmeke2017audioset} and sound event detection~\autocite{desed2020dataset} tasks.
Therefore, nested \textit{K-fold} cross-validation with stratified class sampling has been used throughout.

We employ 5-fold and 10-fold cross validation for the \ac{elev} dataset and \ac{ldc} datasets, respectively.

\subsection{Data pre-processing}
The following section describes how the raw audio was pre-processed before it was used for model training and evaluation.

\subsubsection{Resampling and input normalisation}
Since some recordings span several minutes, each recording was first divided into shorter intervals, while ensuring no call was split.
Furthermore, all audio recordings are resampled to a common \qty{16}{\kilo\hertz} sampling rate.
The \textit{LDC} dataset has two audio channels, the first containing the animal sound recording and the second containing voice memorandums with field notes made during the recording.
Only the first channel was used in our experiments.
The \textit{EV} dataset also has two audio channels, in this case a stereo recording.
These channels were averaged to produce a single channel recording.
Finally, all recordings were normalised to have zero mean and a peak amplitude of \qty{-1}{\dB}.

\subsubsection{Feature extraction}
\label{sec:feature-extraction}
Spectral-temporal feature representations, in the form of mel-frequency spectra and mel-frequency ceptra were computed from the resampled audio signal, using \qty{25}{\milli\second} frames, a stride of \qty{10}{\milli\second} and a Hamming window function.
While these values are common in speech processing tasks, their efficacy has yet to be shown for elephant vocalisations.
We experimentally evaluate different frame lengths and inter-frame stride, for a set of shallow classification models.
However, some of the pre-trained deep-neural models use a standardised frame length and skip.
In such cases this configuration was maintained across models, to allow the direct comparison between model architectures, and to allow for the use of transfer learning.

For mel spectrograms, a frame-wise spectral representation is computed by computing a 1024-point \ac{dft} from a zero-padded and windowed frame.

The magnitude of the complex-valued discrete spectrum is squared to obtain the real-valued power spectrum.
The power spectrum is mapped to the mel-frequency scale using a bank of $N_m$ log-spaced triangular filters.

The resulting mel-scale spectrogram can be further transformed to a mel-frequency cepstrum representation by taking the logarithm of the mel-scale power spectrum and applying the discrete cosine transform.
Of the resulting $N_m$ \acp{mfcc}, the first $N_c$ coefficients are retained.
\Acp{mfcc} features were only evaluated for shallow classification models.

\section{Call detection, endpointing and classification}
\label{sec:call-det-end-class}

\emph{Call detection} is the determination of whether an elephant vocalisation, of any type, occurs in a certain audio signal.
These audio signals can be highly \emph{polyphonic}, in other words, contain multiple other audio sources other than elephant vocalisations, as opposed to isolated elephant vocalisations.
These additional sources may be biological~(e.g. birds) or mechanical~(e.g. vehicles).

\emph{Call endpointing}, is the determination of the beginning and end of a call, within an audio signal.
When call detection is performed for every frame in an audio signal, endpointing is implicitly also performed at the resolution of the frame stride.

\emph{Call classification}, is the identification of the type of vocalisation~(\texttt{rumble}, \texttt{trumpet}, etc.) associated with the call isolated during endpointing~(segment).
We define a \emph{segment}, to be an interval of audio that has already been endpointed to contain a particular call.
This is in general a multi-class multi-label classification problem, because there are more than two call types~(multi-class) and more than one call may occur simultaneously~(multi-label).

In the above description, call detection a binary classification problem (\texttt{call} / \texttt{no-call}).
It is however also sometimes considered as a multi-class multi-label classification problem, where not only the presence and location in time of the call is determined, but also its identity~(\texttt{rumble}, \texttt{trumpet}, etc).

From here on we shall refer to multi-class multi-label call detection simply as multi-label call detection.

\begin{figure}[htb]
  \centering
  \begin{subfigure}{0.45\linewidth}
    \centering
    \includegraphics[width=\textwidth, page=2]{background/melspec_annotation}
    \caption{Call detection --- frame level task.}
    \label{fig:background:melspec:det}
  \end{subfigure}
  \hspace{0.5cm}
  \begin{subfigure}{0.45\linewidth}
    \centering
    \includegraphics[width=\textwidth, page=1]{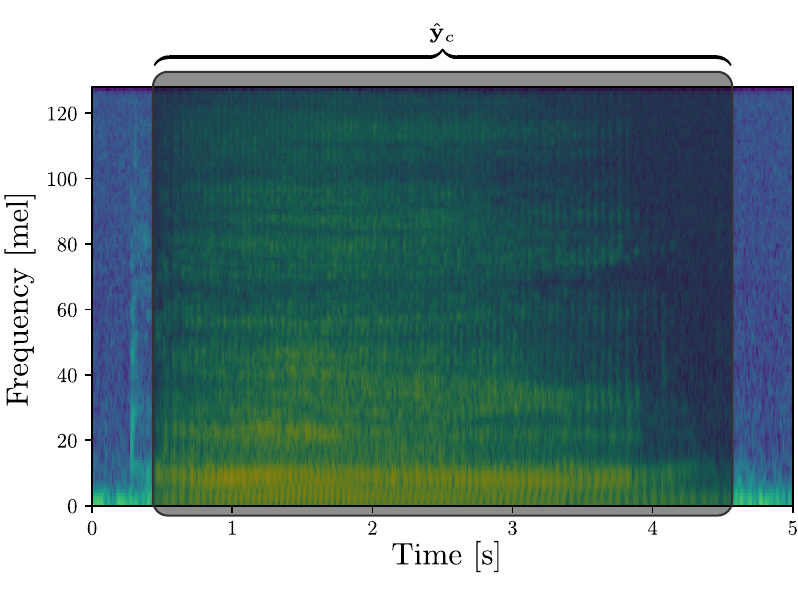}
    \caption{Call classification -- segment level task.}
    \label{fig:background:melspec:class}
  \end{subfigure}
  \caption{
    Mel-spectrogram feature representation of an elephant call used to illustrate (\subref{fig:background:melspec:det}) call detection and (\subref{fig:background:melspec:class}) call classification.
    On the left, $\hat{\mathbf{y}}^{(i)}_{d}$ denotes the classifier output for the $i$-th frame in a sequence of $N$ frames extracted from one recording.
    The shaded area indicates the frames for which a positive detection decision was made.
    On the right, $\hat{\mathbf{y}}_{c}$ denotes the multi-label classifier output for a single elephant call, already endpointed~(shaded).
  }
  \label{fig:background:melspec}
\end{figure}

\Cref{fig:background:melspec} illustrates the difference between call classification and call detection.

\section{Experimental Setup} \label{sec:experiments}
In this section we describe the procedure used to train and evaluate the classifiers described in \cref{sec:background}.

For binary call detection, we follow a similar approach to \textcite{zeppelzauer2013eledetmethod}.
Unless otherwise stated, we apply each model to a context window (consisting of one or more spectral feature vectors) and compare a single multi-label output.
Subsequently, the context window is shifted in time, and the next model output is computed.

For call classification, an oracle model is assumed to have segmented the call, meaning that endpoints are obtained from the ground-truth annotations.
The classifier is then applied to this segment of the audio signal.
This results in several per-frame multi-label classifications for a single call segment.
We compute the average\footnotemark of these per-frame classifications to determine the result for the segment.
\footnotetext{
  This differs from \textcite{zeppelzauer2013eledetmethod}, who used the maximum classifier output probability instead.
}
For model architectures that support multi-label output, such as random forest and \ac{mlp} with a sigmoid output activation, no further modifications to the models are required.
In the case of model architectures that do not, such as logistic regression, we train $C$ binary classifiers, one for each of the $C$ classes.
In this work we regard the multi-label call classification to be a down stream task of call detection and endpointing.

Thus, results presented for call classification are assumed to be down stream from an oracle call detection system.
We evaluate call detection using both a binary~(e.g., \texttt{call} / \texttt{no-call}) and multi-label~(any of \texttt{rumble}, \texttt{trumpet}, etc.) schemes, while we always regard the task of call classification in a multi-label scenario.

\subsection{Call detection} \label{sec:call-event-det}
We perform call detection using two different strategies: \emph{per-frame classification} and \emph{one-to-one sequence classification}, both illustrated in \cref{fig:classification-arch}.
The first strategy takes as input a context window $\mathbf{X}^{{(i)}}$ composed of one or more sequential spectral features vectors~$X^{(i-\frac{w}{2})} \ldots X^{(i+\frac{w}{2})}$.

The spectral features themselves are computed as described in \cref{sec:feature-extraction}.
The classifier produces a single classification output associated in time with the $i$-th~(centremost) frame of the input context window.
Experimentally, we evaluate different input context window lengths, while keeping the per-frame classification stride length fixed, allowing for comparable results.
The second strategy takes as input a slightly different context window~$\mathbf{X}^{(i,k)}$ consisting of frames $i$ to $k$ of the input signal and produces a sequence of classification outputs of the same length.

\subsubsection{Target classification rate}

We have fixed the output per-frame classification stride to \qty{100}{\milli\second} with a \qty{50}{\percent} frame overlap for all experiments.
Hence, one classification is made every \qty{100}{\milli\second}.
When scoring, the start and end of recordings are truncated according to the model with the longest context length to allow for the direct comparison of results between models.

The target labels are obtained from the annotation files using the same classification stride of \qty{100}{\milli\second}.
An annotation label must apply to at least \qty{20}{\percent}~(\qty{20}{\milli\second}) of the classification frame to be considered a valid label for that frame.
The target label is produced irrespective of the length of the input context window, which in some experimental evaluations is longer than the target label frame length.
In such cases, the classifier is tasked with classifying the centre \qty{100}{\milli\second} of the input context window.

\subsubsection{Shallow classification models}
\begin{figure}
  \centering
  \begin{subfigure}{0.45\linewidth}
    \includegraphics[width=\textwidth, page=1]{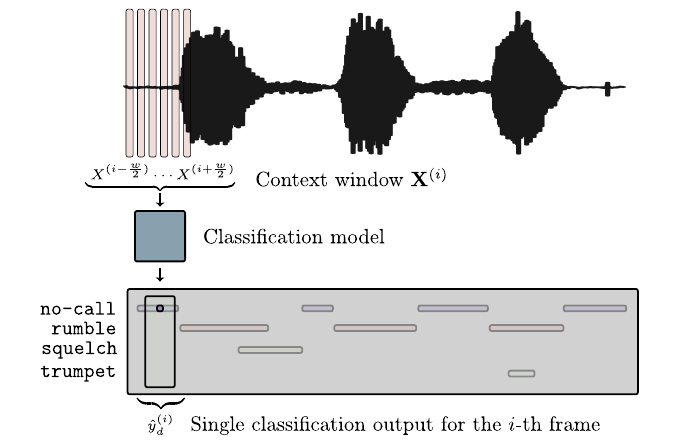}
    \caption{Call detection by means of per-frame classification.}
    \vspace{\baselineskip}
    \label{fig:call-event:framewise}
  \end{subfigure}
  \hspace{0.5cm}
  \begin{subfigure}{0.45\linewidth}
    \includegraphics[width=\textwidth, page=2]{model/call_event_detection}
    \caption{Call detection by means of one-to-one sequence classification.}
    \label{fig:call-event:sequence}
  \end{subfigure}
  \caption{
    Illustration of the two strategies followed for call detection, described in \cref{sec:call-event-det}.
    On the left, $\hat{y}_{d}^{{(i)}}$ denotes the classifier output for the $i$-th~(centre) input frame, given the input context window~$\mathbf{X}^{{(i)}}$ consisting of $w$ consecutive spectral features~$X^{{(i-\frac{w}{2})}} \ldots X^{(i+\frac{w}{2})}$.
    On the right, $\left[ \hat{y}^{(i)}_{d}\ldots \hat{y}^{(k)}_{d}\right]$ denotes the classification sequence produced by the classifier for frames $i$ to $k$, given an extended context window~$\mathbf{X}^{(i,k)}$.
  }
  \label{fig:call-event}
\end{figure}

As shown in \cref{fig:call-event:framewise}, shallow classification models produce a single output estimate per input context window~(per-frame approach).
This window is shifted in time by classification frame stride, after which the next classification is made.

This results in a sequence of classification results for a given audio signal.

We evaluate the shallow models both when using a single spectral frame as input, and also when providing several concatenated spectral frames~(context window).
The length of the context window~(number of concatenated frames) is treated as an experimental hyperparameter.
Such concatenation can lead to high-dimensional input vectors, which may in turn lead to overfitting, or poor convergence during training.
We therefore additionally evaluate \ac{pca} as a dimensionality reduction step.

This dimensionality reduction, as well as cepstral mean and variance normalisation, are also considered as experimental hyperparameters.

\subsubsection{SVM probability calibration}
Certain models, such as the \ac{svm}, do not produce a class membership probability as output.
Instead, these models provide a classification score.
Probability calibration, in the context of classification models, is the process of using an additional classifier or regression model to map these classification scores to a class probability.

We considered two methods of achieving this.

The first method is Platt scaling~\autocite{platt1999scaling}, which adds a logistic regression classifier to the output of the \ac{svm} to produce the desired target label probability.
The second method is isotonic regression~\autocite{zadozny2001isotonic,zadozny2002isotonic}, which fits a piecewise-constant non-decreasing function, to the output of the decision function.
Isotonic regression was chosen as it has shown to offer better performance than Platt scaling, especially when sufficient training data is available~\autocite{mizil2005classcal}.

\subsubsection{Kernel approximation}
In the case of \ac{svm} classifiers, we experimentally evaluate the use of different non-linear kernel functions.
In order to compute the exact kernel function requires on the order of $\mathcal{O}\bigl(N_{s}^{3}\bigr)$ computations.
This cubic scaling in the number of input samples (frames) is computationally intractable for the datasets used in this work.
Thus, we consider two methods for approximating the kernel function, the first is using \textit{Nystr\"om} method~\autocite{williams2000nystroem} and the second is \textit{Random Kitchen Sinks} method~\autocite{rahimi2008randkitchensink}.

The particular kernel function and kernel approximation methods used are left as an experimental hyperparameter.
However, we evaluate both the radial basis function and a third degree polynomial as a basis for the kernel function.

\begin{figure}
  \centering
  \begin{subfigure}{0.45\linewidth}
    \centering
    \includegraphics[page=1]{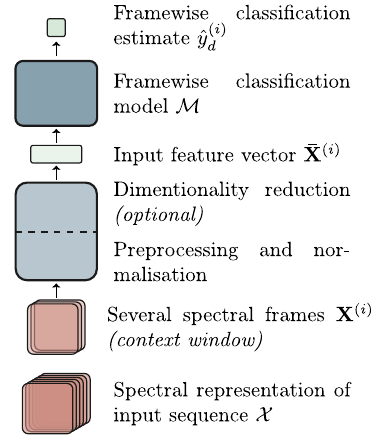}
    \caption{Call detection by means of per-frame classification.}
    \label{fig:classification-arch:framewise}
  \end{subfigure}
  \hspace{0.5cm}
  \begin{subfigure}{0.45\linewidth}
    \centering
    \includegraphics[page=2]{model/model_overview}
    \caption{Call detection by means of one-to-one sequence classification.}
    \label{fig:classification-arch:seq2seq}
  \end{subfigure}
  \caption{
    Model overview of classification models used for call detection.
    Only a single input sequence is shown, and batch processing is omitted from the illustration.
  }
  \label{fig:classification-arch}
\end{figure}
\subsubsection{\Acl{cnn}}
Beyond shallow classification models, we also evaluate a series of deep convolutional classifier architectures, including AlexNet~\autocite{krizhevsky2012alexnet}, ResNet~\autocite{he2016resnet}, and VGGNet~\autocite{liu2015vggnet}.
Unless otherwise specified, all deep architectures were trained using the \textit{Adam}~\autocite{kingma2014adam} optimiser with $\beta_{1}=0.9$ and $\beta_{2}=0.999$ with binary cross entropy as the loss function.
The initial learning rate of the optimiser is considered a hyperparameter.

To ensure that the model is adequately trained, we monitor both the training and the development loss for convergence, or divergence.
If either loss diverges (increases) for more than three epochs or converges (remains unchanged within a threshold), the training process is terminated.
In the case of divergence, we rewind the model parameters to the previous state with the lowest developmental loss.
For AlexNet and ResNet architectures, we also test whether transfer learning on an out-of-domain image dataset (e.g. ImageNet) improves performance compared to random initialisation of the model~\autocite{deng2009imagenet}.

Like the shallow classifiers, the convolutional models take a temporal-spectral feature representation~(mel-spectrogram) as input.
The exact configuration of this mel-spectrogram is discussed in \cref{sec:feature-extraction}.
Context window lengths in the range \qtyrange{0.5}{3}{\second} are considered experimentally, but in all cases a single classification output associated with the centre \qty{100}{\milli\second} frame of the context window is produced.
Hence, like the shallow architecture, we use \acp{cnn} to achieve call detection by means of per-frame classification, as shown in \cref{fig:classification-arch:framewise}.

We evaluate the standard model configuration for AlexNet~(\qty{61.1}{\mega\unitless} params.), while for VGGNet we evaluate four variations of different size, the largest of which contains \qty{143.6}{\mega\unitless} parameters.
While VGGNet was considered, it lead to no noticeable performance increase over AlexNet, and is thus omitted from the results.

We consider the five standard ResNet configuration with the smallest, \textit{ResNet-18}, containing \qty{11.7}{\mega\unitless} parameters and the largest, \textit{ResNet-152}, containing \qty{60.2}{\mega\unitless} parameters.

\subsubsection{Transformer-encoder}
Finally, we evaluate the transformer-encoder as a one-to-one sequence classification model for call detection, as shown in \cref{fig:classification-arch:seq2seq}.
Unlike the classification models described in the previous sections which produce a single classification estimate per input sequence, these architectures produce as many classification outputs as there are input sequence tokens~(frames).
Since this approach can produce multiple per-frame classifications within a single inference or training step, it can take advantage of parallel computation.
Such models have the ability to model temporal dependency and make estimations based on prior, and in some cases future, observations.

\Cref{fig:classification-arch} provides an illustration of such sequence classification models, compared to typical per-frame classification models.

We choose the \ac{ast} one-to-one sequence classification model for experimentation, described in \cref{sec:background:ast}.
We omit the learned classification token that is usuall prepended to the input sequence and instead we train a \ac{mlp} prediction head directly on the output representation of the transformer model~\autocite{chen2022beats}.
As the model produces 2D spectral-temporal representation as output, we average the spectral features to obtain a single temporal representation per time step.
This variant of the model is referred to as \gls{ast-seq}.

For encoder-only transformer models, there exists a one-to-one correspondence between the number of input and output tokens.
Specifically, for an input sequence of containing $N$ tokens, the model will produce $N$ outputs~(\cref{sec:background:ast}).
In the case of the \ac{ast} architecture, each input sequence token consist of 16 spectral frames.
Thus, a classification estimate is produced approximately once every \qty{160}{\milli\second}, which does not exactly match the classification rate of one classification per \qty{100}{\milli\second} used in our other experiments.
To address this discrepancy, the classification output is resampled using linear interpolation to the correct frame rate.
However, this resampling process is not performed during model training.

We experimentally evaluate the effectiveness of training the \ac{ast} model from a random initialisation, as well as pre-training the model using the self-distillation scheme presented by \textcite{chen2022beats} on AudioSet~\autocite{gemmeke2017audioset}.
During fine-tuning we employ a training regime, similar to that proposed by \textcite{google2022freezebackbone}, in order to preserve semantic information obtained during pre-training.
For the first \qty{10}{epochs} of training, the backbone feature extraction neural network is fixed~(no gradient update is performed).
After this the backbone network is updated with a learning rate that is \qty{1}{\percent} of the \ac{mlp} prediction head learning rate.
The backbone learning rate is then increased over the next two epochs to match the overall learning rate.
This technique is common practice in fine-tuning models as it reduces the chance of losing transferable semantic information learned during pre-training and has been showed to improve performance on task~\autocite{google2022freezebackbone}.
The model's parameters are optimised using the \textit{AdamW}~\autocite{losh2017adamw} optimiser with $\beta_{1}=0.9$ and $\beta_{2}=0.98$, and a weight decay factor of 0.01.

In addition to \gls{ast-seq}, we also evaluate the standard architecture as proposed by the \autocite{gong2021ast}, which uses the learned classification token \texttt{[CLS]} to produce a per-frame classification.
This variant of the model is referred to as \gls{ast-lab}.

\subsection{Call endpointing}
Call endpointing refers to the specific task of locating the precise beginning and end of an elephant vocalisation within a continuous audio recording.

By accurately identifying the start and end times of elephant calls, the efficiency and accuracy of subsequent data processing tasks is enhanced.

Endpointing is performed implicitly in this work, through per-frame binary call detection as described above.
In order to achieve endpointing, a threshold~(e.g. $\theta = 0.5$) is applied to each of the per-frame call detection probabilities to classify each frame as a positive~(\texttt{call}) or negative~(\texttt{not-call}).
The start of a segment~(new call) is identified by a negative call detection frame followed by a positive frame~(rising edge), the end of segment~(call) is identified by a positive frame followed by a negative frame~(falling edge).

\subsection{Call classification}
Call classification aims to determine which call or calls are present in a presented audio segment.
The task of call classification seeks to determine which of the major or subcall types a particular audio sequence belongs to.
In sound classification tasks it is typically assumed that there is a single dominant acoustic source present and that the audio is already endpointed --- i.e.\ the start and the end of the call are also the start and end of the presented audio segment.
This implies that there is no temporal context indicating for where in the overall recording the call occurs.

In order to perform call classification we used the same models used for call detection and therefore also the same hyperparameters.
The model produces several per-frame classification estimates for each call segment, and  call classification is achieved by averaging the per-frame call detection probabilities over the endpointed segment.
This average is considered the output of the call classifier models.

As before, the target label for each segment is obtained from the annotations.
However, a single multi-label target output is assigned to each call segment to be classified.
In cases where neighbouring calls overlap with a given segment, the neighbouring call must occupy at least \qty{50}{\percent} of the call segment to be considered present.
This does mean that the interval of overlap between such calls is evaluated twice, once for the first segment and again for the succeeding segment.

\subsection{Performance evaluation metrics}
Several performance metrics will be used to assess the ability of the models to perform call detection, endpointing and classification, as described in the previous sections.

\subsubsection{Dealing with class imbalance}
The datasets utilised in this study exhibit considerable class imbalance, an issue that is prevalent in many animal vocalisation studies (\cref{sec:target-label}).
This disparity may distort performance assessments based on certain metrics.
To mitigate the influence of class imbalance on multi-label classification models, we implement two strategies: one-vs-one scoring and macro-averaging across all classes.

\textbf{One-vs-rest classification.}

In multiclass scenarios, binary classification metrics can be applied by considering comparisons between a target class (positive) and all the other classes (negative).
This approach produces a set of scores for each classification category and is referred to as one-vs-rest classification.
However, one-vs-rest metrics remain vulnerable to class imbalance~\autocite{bishop2006mlpattern,lorena2008combmulticlass}.

\textbf{One-vs-one classification.}

A variation of one-vs-rest classifiers is the one-vs-one approach, in which a single target class is contrasted with a single different class, for instance, \textit{rumble}-vs-\textit{trumpet}.
One-vs-one scores demonstrate resilience to class imbalance when employing robust metrics.
Since all classifiers presented in this paper include an explicit no-call class, we evaluate each class as a binary classifier versus the no-call class (e.g. \textit{rumble}-vs-\textit{no-call}).

\textbf{Multi-class macro-averaging.}
Certain classification metrics, such as classification accuracy, may suffer from misleading results if the dataset is unbalanced.

One way of overcoming this is to compute the metric for each respective class~(\emph{one-vs-rest} or \emph{one-vs-one}) and then compute the average.
Since each individual metric is normalised within its class, this avoids skew.
Using one-vs-one, as individual metrics has been shown to be more robust to class imbalance~\autocite{bishop2006mlpattern,lorena2008combmulticlass}..

\subsubsection{Classification metrics}
Call classification performance was assessed using the binary metrics: sensitivity, specificity, precision and recall, and derived quantities.
All performance metrics reported are calculated separately for each outer cross-validation fold and then averaged.
We consider if a particular call type is present as a the positive class and vice versa.
We treat each separate class (call type) as its own binary classification problem~(multi-label).

\textbf{Specificity and sensitivity.}
Specificity measures the true negative rate, i.e. the proportion of true negatives among all actual negatives, given by:
\begin{equation}
    Specificity = \frac{TN}{TN+FP} = \frac{TN}{N} = 1 - FPR
\end{equation}
Sensitivity, also known as recall, measures the proportion of true positives among all actual positives, given by:
\begin{equation}
    Sensitivity = Recall = \frac{TP}{TP+FN} = \frac{TP}{P} = TPR
\end{equation}
A high specificity implies a low false positive rate~($FPR$), while a high sensitivity implies a high true positive rate~($TPR$).

\textbf{\Acl{roc}.}
Specificity and sensitivity reflect classification performance at a single operating point~(i.e.\ decision threshold).
Increasing the decision threshold will result in an increase in the system's sensitivity, at the cost of a decrease in its specificity, and vice versa.

To visualise the trade-off between system specificity and sensitivity is the \ac{roc}.
A \ac{roc} curve is drawn by plotting a locus of $TPR$ against $FPR$ for a series of decision thresholds.

\textbf{\Acs{auc roc}.}
The area under the \ac{roc} curve~(\acs{auc roc}) provides a single-figure indication of classifier performance across all decision thresholds.
An \ac{auc roc} score of \num{1.0} indicates a perfect classier, whereas a score of \num{0.5} indications the classification performance achieved by guessing the prior probability~(i.e.\ a random classifier).

\textbf{Precision and recall.}
As alternatives to specificity and sensitivity, two other commonly used metrics are precision and recall.
Recall is equivalent to sensitivity.
Precision is the ratio of true positive~($TP$) classifications to the total number of positive classifications~($PP$) made by the model.
It measures the proportion of correctly identified positive instances (true positives) out of all the instances labelled as positive (both true and false positives).
For a binary classifier, precision is defined as:
\begin{equation}
    Precision = \frac{TP}{TP+FP} = \frac{TP}{PP}
\end{equation}

\textbf{Precision-recall curve.}
The \emph{precision-recall curve} is the analogue of the \ac{roc} curve.
In this case we vary the decision threshold and obtain a locus of precision and recall values.
Unlike the \ac{roc} curve, the precision-recall curve is not monotonically increasing.

\textbf{Average precision.}
For a binary classifier, the area underneath the precision-recall curve is defined as the \ac{ap}.
If the score is evaluated in a multi-class scenario, the average over all classes is computed and is referred to as the \ac{map}.

\subsubsection{Detection and endpointing metrics}
For the task of call detection, models were evaluated based on per-frame coverage and purity as well as the Jaccard index.
These metrics are accepted practice in the field of acoustic segmentation~\autocite{kemp2000segmetrics}.

\textbf{Boundary precision and recall.}

When applied to endpointing, the metrics precision and recall can be computed based on the number of correctly detected boundaries.
In this case, recall represents the number of boundaries that are correctly detected as a fraction of all postulated boundaries, while precision reflects the fraction of postulated boundaries that are indeed correct.
Using precision and recall in this manner requires the introduction of a boundary tolerance parameter.
A postulated boundary that falls within this margin of a true boundary will be regarded as a~$TP$.
This parameter must be kept constant between experiments to ensure that the associated metrics are directly comparable.

\textbf{Coverage and purity.}
Coverage and purity are popular metrics in the speaker diarisation field.
The metric is used to evaluate how well a postulated segment is representative of the ground truth segment.
These metrics pose as an alternative to boundary precision and recall, and benefit from not requiring an explicit boundary tolerance.
However, due to the discrete per-frame evaluation of the metric, the interval between successive classification frames~(frame stride) imposes an intrinsic scoring resolution.

For a per-frame binary classifier, coverage is defined as the ratio between the correctly classified positive frames and the number of positive samples.
Therefore, coverage is merely per-frame recall, and is indicative of how well the classifier is able to detect calls.
The associated metric of per-frame precision is known as  purity, and reflects the fraction of positively classified frames that are indeed correct.
Purity provides insight into how precise the positive classifications are.
The combination of these metrics aim to measure a models' tendency to over- or under-segment a sequence.

Over-segmentation is the result of a model that over-eagerly identifies boundaries in a sequence, thus leading to excessive fragmentation.
Such models generally exhibit an increased number of false-negative classification and thus lead a decrease in coverage, but also tend to exhibit high purity.
On the other hand, under-segmentation occurs when decisions are overly conservative, leading to an increase in false-positive classifications, and thus a reduction in purity while maintaining a high degree of coverage.

\begin{figure}
  \centering
  \includegraphics{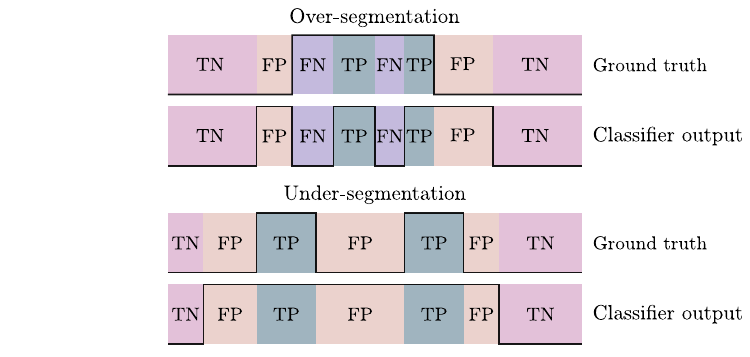}
  \caption{
    Illustration of over- and under-segmentation.
    Note, while the illustration depicts a continuous segmentation process, in this work we perform the segmentation using per-frame detection and thus it is intrinsically discrete.
  }
  \label{fig:segmentation}
\end{figure}

The ideal segmentation model would have both high coverage while maintaining a high purity score.
\Cref{fig:segmentation} illustrates both over- and under-segmentation.

\textbf{Jaccard index.}
The final metric of we describe is the Jaccard index.
While predominantly used in image segmentation tasks, where it is known as \ac{iou}, or numerical set similarity, it has seen application in one dimensional segmentation tasks.
The Jaccard index of two sets is defined as follows:
\begin{equation}
    J(y, \hat{y} \mid y_c = \hat{y}_c) = \frac{\mid y \cap \hat{y} \mid}{\mid y \cup \hat{y} \mid}
\end{equation}
where $y$ and $\hat{y}$ are the ground truth and the classifier output, given that the class is the same~($y_c = \hat{y}_c$) while $\cap$ and $\cup$ are the intersection and union between two sets, respectively, and $\mid \cdot \mid$ is the cardinality of the set (in our particular case the duration in frames).
For a per-frame binary classifier, the Jaccard index is given by:
\begin{equation}
  J = \frac{TP}{TP+FN+FP}
\end{equation}
The Jaccard index is a measure of the system's segmentation localisation ability.

The Jaccard index is bound between $0.0$ and $1.0$, where $1.0$ indicates a perfect segmentation model.

The Jaccard index accounts for both the coverage and purity of a segmentation model.
One can thus not distinguish purely based on the Jaccard index whether a model is suffering from over- or under-segmentation.
However, the Jaccard index encapsulates overall segmentation performance as a single number.

\subsection{Hyperparameter selection}
\label{sec:exp-setup:hparam}
In this section we review our findings regarding the observed hyperparameters selections for our experiments.
The hyperparameters have been selected by employing nested cross-validation, described in \cref{sec:kfold}.
The best model is determined based on the model hyperparameter configuration with the lowest loss across for the inner cross-validation folds.
This results in $K-1$ experimental configurations, from which the most common hyperparameter choices used to re-train the model on all inner folds~(i.e.\ no development set).
This final model is evaluated on the respective outer-fold.
This is repeated for all outer-folds.
Finally, the $K$ outer-fold scores are averaged to obtain the overall performance.

\subsubsection{Feature extraction parameters}
In the following section we discuss the spectral extraction feature selection, using a \ac{lr} classifier for the task of per-frame binary call detection.
For this architecture we tested a broad range of \ac{stft}, mel-spectrum and \ac{mfcc} configurations.

Initially we experiment by using a single \ac{mfcc} feature as input to the \ac{lr} classier, we denote this model as \gls{lr-frame}.
Our findings indicate that, when only a single \ac{mfcc} frame is taken as input, increasing the \ac{stft} frame length from \qty{25}{\milli\second}~(typical in speech applications) to \qty{100}{\milli\second} consistently enhanced performance.

Although there is a discernible pattern favouring a minimal number of mel triangular filters (64 mel filters), the quantity of \acp{mfcc} retained does not appear to negatively impact performance.
These higher order \acp{mfcc} are known to encode speaker pitch information when applied to speech signals.
Further subsequent analysis of these coefficients revealed reduced statistical variance of these coefficients compared to the lower order \acp{mfcc}.
We thus believe that, while the use of these higher order cepstral coefficients did not strongly impact classifier performance, their utility when processing elephant vocalisations still remains uncertain.

Subsequently, we considered using multiple neighbouring \ac{mfcc} features~(context window) as input to the classifier, and denote this classifier simply as \acs{lr} in our results.
In contrast to \gls{lr-frame}, our experiments show that, when using a context window, reducing the \ac{stft} frame length to \qty{500}{\milli\second} led to improvements for both datasets.
The choice of a smaller number of mel filters did not generalise to the context window input experiments.
Instead, we found that a larger number of mel filters led to better performance.
Furthermore, retaining only the lower order~(31) \acp{mfcc} lead to further improvements.

As noted before, the higher order \acp{mfcc} exhibited reduced variance compared to their lower order counterparts.
Further experimentation employed \ac{pca} as a dimensionality reduction instead of discarding the higher order \acp{mfcc}.
The number of principal components to retain was chosen such that the explained variance is \qty{95}{\percent} of the total features (7~components).
The choice of \ac{pca} dimensionality reduction technique stemmed for the observed reduced variance in the higher order \acp{mfcc}.
Employing \ac{pca} as a preprocessing step outperformed the n\"iave approach of merely discarding the higher order coefficients.

The multi-frame \ac{lr} models exhibited substantially better performance than their single frame counterparts.

\subsubsection{\Acl{mlp}}
From our experiments we found that, over both datasets, the \ac{mlp} architecture and the layerwise dropout probability lead to the most substantial performance enhancement on the development sets.
A choice ranging between \qtyrange{10}{15}{\percent}, lead to consistent perforamance improvement.

Contrary to expectations, an encoder structure with decreasing hidden layer dimensions as network depth increased did not result in the best performance.
Instead, maintaining a consistent hidden layer dimension proved to be the most effective approach.

While deeper \acp{mlp}~(more than five layers) showed improved performance on certain development folds for the \ac{elev} dataset, shallower structures, comprising of two to four layers (less than 1 million parameters) consistently outperformed the deeper \ac{mlp} models.
However, this trend did not generalize to the \ac{ldc} dataset, where deeper models consisting of five to seven layers~(containing up to 10 million parameters) exhibited superior performance.
This can be attributed to the difference in dataset sizes.

The activation function used in the hidden layers had minimal impact on overall performance.
Furthermore, batch normalisation adversely affected model performance for both datasets.

\subsubsection{\Acl{svm}}
In contrast to \ac{lr} and \acp{mlp}, no noticeable performance improvement was achieved by optimising the context window, or \ac{mfcc} configuration when using a \ac{svm} classifier.
The range of \ac{mfcc} configurations was not as extensive as for \acs{lr}[-frame], only evaluating the top performing configurations.
However, the choice of a long context window with a \ac{pca} dimensionality reduction step to \qty{5}{components}~(\qty{92.5}{\percent} explained variance) consistently lead to better performance.
The choice of kernel approximation function had the most significant impact on overall performance.
The Nystr\"om method lead to a reduction of \qty{50}{\percent} on the developmental loss, compared to the \textit{Random Kitchen Sinks} method, matching the exact kernel performance on the \ac{elev} dataset.
The \ac{rbf} was the best overall performing basis function.

\subsubsection{ResNet}
Out of the standard ResNet configurations~(\textit{ResNet-18}, \textit{ResNet-34}, \textit{ResNet-50}, \textit{ResNet-101} and \textit{ResNet-152}) evaluated, \textit{ResNet-101} performed the best when evaluated on the \ac{ldc} dataset.
While, the best configuration for the \ac{elev} dataset was \textit{ResNet-18}.
In both cases a context window of \qty{2.5}{seconds}, performed the best overall.
There was no clear trend on the choice of learning rate or batch size for training the ResNet models.

\subsubsection{AlexNet}
As expected, \acp{cnn} see improved performance with longer context windows.
When using AlexNet, the best performance was observed with a context window of \qty{2.5}{seconds}, the same as seen for the ResNet architecture.

\subsubsection{In-domain pre-training}
The \ac{ast} model was evaluated using model weights that had been pre-trained on AudioSet as well as weights that had a random initialisation.
When training from the random initialisation, the model would not generalise well and resulted in poor developmentset performance.
However, when fine-tuning from a pre-trained model, the \ac{ast} model outperformed all other architectures.
Utilising a pre-trained model weights required lowering the initial learning rate of the optimiser.

\subsubsection{Out-of-domain pre-training}

The impact of out-of-domain pre-training was assessed for both ResNet and AlexNet architectures.
It was observed that pre-training these models on out-of-domain ImageNet data resulted in marginal improvements in performance compared to random initialisation.
In the case of ResNet, this approach facilitated the use of larger models.
However, these did not yield any further enhancements in classification performance.
Consequently, employing a smaller ResNet model may be more advantageous due to its better computational efficiency.
Pre-training AlexNet on ImageNet led to marginal performance improvements.
Remarkably, the optimal model parameters were obtained after just one epoch of training when initialised with pre-trained weights, thereby significantly reducing computation time.
Notably, in both ResNet and AlexNet models, a reduction in the initial learning rate by two orders of magnitude was necessary, compared to training from random initialisation, to achieve generalisable performance.

\section{Results}
\label{sec:results}
The following section described the results obtained for the experiments set out in \cref{sec:experiments}.

\subsection{Call detection and endpointing}
\begin{figure}
  \begin{subfigure}{0.5\linewidth}
    \centering
    \caption{
      \Acs{elev} dataset
    }
    \label{fig:roc:frame:bin:elev}
    \includegraphics[scale=0.75]{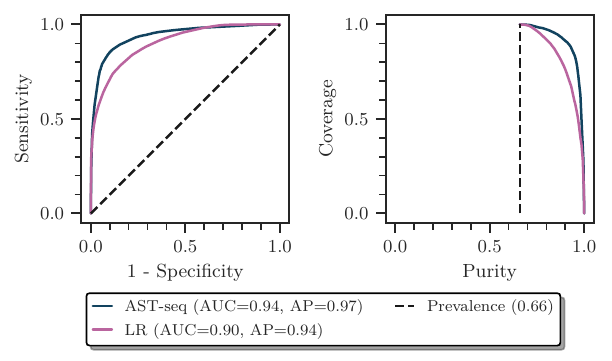}
  \end{subfigure}
  \begin{subfigure}{0.5\linewidth}
    \centering
    \caption{
      \Acs{ldc} dataset
    }
    \label{fig:roc:frame:bin:ldc}
    \includegraphics[scale=0.75]{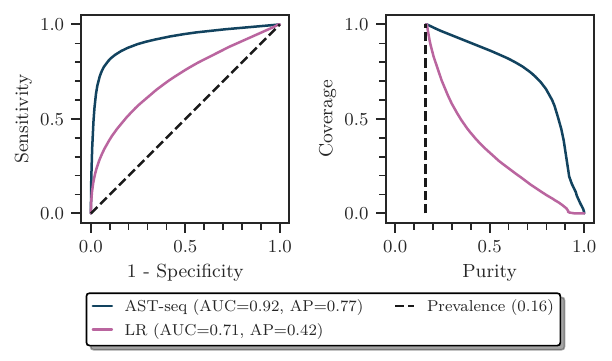}
  \end{subfigure}
  \caption{
    \Acl{roc} and coverage-purity curves, for the task of per-frame elephant call detection, both \acf{elev} and \acf{ldc} datasets.
    The baseline model~(\ac{lr}) and the top performing model~(\gls{ast-seq}) are shown.
    The prevalence, which is the performance achieved when making random decisions according to the prior.
  }
  \label{fig:roc:frame:bin}

\centering
\captionof{table}{
  Per-frame elephant binary call detection results for the \acf{elev} and the \acf{ldc} datasets for all considered models.
  The reported metrics are averages over the $K$ outer folds.
  The associated standard deviation is given in parentheses.
}
\label{tab:results:detection:binary}
\label{tab:results:detection}
\label{tab:classification-results}

\sisetup{
  detect-weight,
  mode=text,
  table-alignment-mode = format,
  table-format = 1.3,
  table-number-alignment = center,
  separate-uncertainty = false,
}

\renewrobustcmd{\bfseries}{\fontseries{b}\selectfont}
\renewrobustcmd{\boldmath}{}

\newrobustcmd{\B}{\bfseries}

\begin{tabular}{lllccccc}
\toprule

Dataset                                   & Model                        &  &    {Purity}     &    {Coverage}   &    {\acs{auc roc}}  &    {\acs{ap}}   &    {Jaccard}    \\ \midrule
\multirow{7}{*}{\acs{elev} ($K=5$)}       & \acs{lr}                     &  &    0.829~(0.04) &    0.798~(0.05) &    0.902~(0.02)     &    0.937~(0.01) &    0.719~(0.04) \\
                                          & \acs{svm}                    &  &    0.853~(0.05) &    0.873~(0.04) &    0.893~(0.02)     &    0.928~(0.02) &    0.757~(0.03) \\
                                          & \acs{xgb}                    &  &    0.857~(0.05) &    0.893~(0.04) &    0.907~(0.02)     &    0.946~(0.02) &    0.776~(0.04) \\
                                          & \acs{mlp}                    &  &    0.869~(0.07) &    0.860~(0.07) &    0.914~(0.03)     &    0.943~(0.03) &    0.754~(0.03) \\
                                          & AlexNet                      &  &    0.873~(0.02) &    0.891~(0.06) &    0.938~(0.02)     &    0.960~(0.01) &    0.787~(0.04) \\
                                          & ResNet                       &  &    0.876~(0.04) &    0.888~(0.02) &    0.933~(0.02)     &    0.957~(0.01) &    0.788~(0.03) \\
                                          & \gls{ast-lab}                &  &    0.829~(0.03) & \B 0.955~(0.02) &    0.938~(0.02)     &    0.958~(0.01) &    0.798~(0.03) \\
                                          & \gls{ast-seq}                &  & \B 0.914~(0.04) &    0.901~(0.02) & \B 0.940~(0.02)     & \B 0.968~(0.01) & \B 0.830~(0.02) \\ \midrule

\multirow{7}{*}{\acs{ldc} ($K=10$)}       & \acs{lr}                     &  &    0.286~(0.03) &    0.602~(0.04) &    0.711~(0.03)     &    0.419~(0.04) &    0.240~(0.02) \\
                                          & \acs{svm}                    &  &    0.634~(0.08) &    0.178~(0.02) &    0.727~(0.02)     &    0.404~(0.05) &    0.161~(0.01) \\
                                          & \acs{xgb}                    &  &    0.664~(0.07) &    0.130~(0.02) &    0.694~(0.02)     &    0.370~(0.05) &    0.121~(0.02) \\
                                          & \acs{mlp}                    &  &    0.665~(0.03) &    0.241~(0.02) &    0.684~(0.03)     &    0.421~(0.03) &    0.201~(0.04) \\
                                          & AlexNet                      &  &    0.691~(0.07) &    0.498~(0.03) &    0.825~(0.03)     &    0.616~(0.04) &    0.403~(0.03)  \\
                                          & ResNet                       &  &    0.633~(0.04) &    0.490~(0.03) &    0.812~(0.03)     &    0.606~(0.04) &    0.382~(0.05) \\
                                          & \gls{ast-lab}                &  & \B 0.768~(0.05) &    0.578~(0.05) &    0.886~(0.04)     &    0.716~(0.04) &    0.504~(0.04) \\
                                          & \gls{ast-seq}                &  & \B 0.768~(0.05) & \B 0.700~(0.04) & \B 0.918~(0.02)     & \B 0.772~(0.06) & \B 0.578~(0.05) \\ \bottomrule
\end{tabular}

\end{figure}

The following experiments assess how well a classification model can perform per-frame elephant call detection and implicit call endpointing, using the procedures described in \cref{sec:call-event-det}.
Model detection performance will be assessed based on the \ac{ap} score, while the endpointing performance will be assessed using the Jaccard index.
\Cref{tab:results:detection} presents experimental results for per-frame elephant call detection.
The table shows than, across both datasets, the \gls{ast-seq} achieves better performance than all other considered models, followed by \ac{cnn}[-based] models~(AlexNet and ResNet) and then the \ac{mlp}.

The \gls{ast}[-based] models clearly outperform the other models, obtaining an \ac{ap} score of \num{0.962} for per-frame call detection.
We speculate this is due to the attention-mechanism used by the transformer-encoder architecture, which allows the model to selectively focus on areas of interest within the extended context window.
Other models have a reduced context window length and also do not have the ability to selectively ignore potential adversarial acoustic events.
While the \gls{ast-seq} model makes more errors per-frame than the \gls{ast-lab} model, informal inspection of the model decisions revealed that these errors tend to occur near the boundaries of the vocalisation segment.
Further inspection revealed that some call boundaries postulated by the model may be more precisely aligned with the vocalisation that the human annotated labels do.
In contrast, the frame errors made by the other models~(in particular the shallow models) are the result of entire vocalisation segments being falsely classified.

From the Jaccard index, we see that the \gls{ast-seq} model produces the best segment alignment with the ground truth annotations, achieving a Jaccard index of \num{0.830}.
However, the high purity and lower coverage indicate that the model tends to over-segment a given audio sequence.
Anecdotally we observe this behaviour particularly in vocalisations that have long tails.
In such cases the model may divide the vocalisation into two or more segments.
In contrast, \gls{ast-lab} tends to under-segment, resulting in missing call boundaries.
We observe this in the lower specificity score, indicative of falsely classifying a segment as a vocalisation.

When computing the average model output over the segment, inserted boundaries have a smaller impact on the final classification than missing boundaries, thus resulting in improved call detection results.
Thus, the \gls{ast-seq} model is the best performing endpointing model, followed by \gls{ast-lab} and ResNet models, achieving a Jaccard index of \num{0.798} and \num{0.788}, respectively.

We observe that the convolutional models, AlexNet and ResNet, are also strong contenders for call detection, matching the transformer models in some metrics.
We speculate that is due to the positional invariant and therefore robustness of the \ac{cnn} features.

The \ac{cnn} models also benefit from increased spectral-temporal resolution in comparison to the \ac{ast} architecture.
The first step in the \ac{ast} architecture is to divide the input spectrogram into patch embeddings, which is a downsampling step.
This step is required to improve the computational efficiency of the model.
However it may result in loss of information that is useful in discerning small variations between features.
Finally, we see that there is a larger performance gap between the \ac{cnn} and transformer architecture when more data is available, as is the case with the \ac{ldc} dataset.

Overall, the deep neural architectures out-performed the shallow classification models in the task of per-frame elephant call detection.
An interesting observation is how well the \ac{svm} classifier performs.
The performance improvement between \ac{svm} and a model of similar computational complexity such as \ac{lr}, can be attributed to the non-linearity present in the kernel function of the \ac{svm} classifier.

A notable architectural observation is that all shallow models saw increased performance when provided with an input context of roughly a second in length and the inclusion of a dimensionality reduction step.
We speculate that this ensures that the models classifications are more robust to the noisy features extracted from each frame.
We did attempt simply averaging the features over time.
However, the dimensionality reduction outperformed this approach, especially when the numerical space was whitened.

\Cref{fig:roc:frame:bin} compares the purity-coverage and \ac{roc} curves for a selection of the models presented in \cref{tab:results:detection}.
From the \ac{elev} purity-coverage curves, we observe that for \qty{95}{\percent} coverage~(i.e.\ successfully identify \qty{95}{\percent} of the true positive frames), \ac{lr} and AlexNet have an almost equal purity~(the identified frames which are indeed true positive) of approximately \qty{81}{\percent}.
This makes \ac{lr} a computationally efficient way to identify frames for further downstream tasks.
From the \ac{roc} curves we see that the \ac{ast-seq} model almost always exhibits better performances.
Furthermore we observe from the prevalence~(the ratio of positive samples to the total number of samples), that the \ac{elev} dataset is more balanced than the \ac{ldc} dataset.
Comparing the \ac{roc} and purity-coverage curves, we see that the performance of both classifiers has decreased for the \ac{ldc} dataset.

\begin{figure}
  \begin{subfigure}{1.0\linewidth}
    \centering
    \caption{
      \Acs{elev} dataset
    }
    \label{fig:proc:frame:call:elev}
    \includegraphics[scale=0.9]{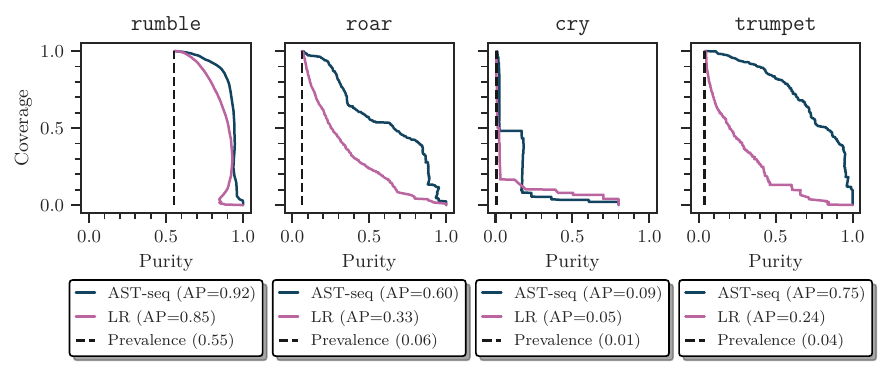}
  \end{subfigure}
  \begin{subfigure}{1.0\linewidth}
    \centering
    \caption{
      \Acs{ldc} dataset
    }
    \label{fig:proc:frame:call:ldc}
    \includegraphics[scale=0.93]{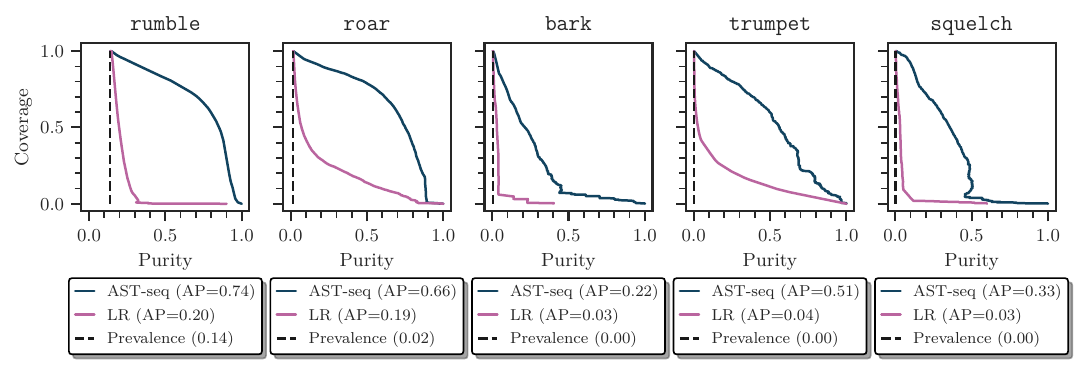}

  \end{subfigure}
  \caption{
    Purity-coverage curves, for the task of per-frame elephant multi-label call detection, both \acf{elev} and \acf{ldc} datasets.
    The baseline model~(\ac{lr}) and the top performing model~(\gls{ast-seq}) are shown.
    The prevalence, which is the performance achieved when making random decisions according to the prior.
  }
  \label{fig:proc:frame:call}

\centering
\captionof{table}{
  Per-frame elephant multi-label call detection results for the \acf{elev} and the \acf{ldc} datasets for all considered models.
  The reported metrics are averages over the $K$ outer folds.
  The associated standard deviation is given in parentheses.
}
\label{tab:results:detection:call}

\sisetup{
  detect-weight,
  mode=text,
  table-alignment-mode = format,
  table-format = 1.3,
  table-number-alignment = center,
  separate-uncertainty = false,
}

\renewrobustcmd{\bfseries}{\fontseries{b}\selectfont}
\renewrobustcmd{\boldmath}{}

\newrobustcmd{\B}{\bfseries}

\begin{tabular}{lllccccc}
\toprule

Dataset                                   & Model                        &  &    {Purity}     &    {Coverage}   &    {\acs{auc roc}}  &    {\acs{map}}  &    {Jaccard}    \\ \midrule
\multirow{7}{*}{\acs{elev} ($K=5$)}       & \acs{lr}                     &  &    0.299~(0.04) & \B 0.511~(0.16) &    0.798~(0.09)     &    0.370~(0.08) &    0.240~(0.04) \\
                                          & \acs{svm}                    &  &    0.433~(0.09) &    0.310~(0.06) &    0.839~(0.08)     &    0.379~(0.06) &    0.249~(0.04) \\
                                          & \acs{xgb}                    &  &    0.507~(0.07) &    0.279~(0.04) &    0.833~(0.08)     &    0.406~(0.05) &    0.240~(0.03) \\
                                          & \acs{mlp}                    &  &    0.397~(0.12) &    0.318~(0.12) &    0.853~(0.07)     &    0.416~(0.07) &    0.258~(0.08) \\
                                          & AlexNet                      &  &    0.515~(0.10) &    0.385~(0.10) &    0.895~(0.07)     &    0.484~(0.09) &    0.314~(0.08) \\
                                          & ResNet                       &  &    0.407~(0.13) &    0.259~(0.05) &    0.847~(0.08)     &    0.385~(0.08) &    0.217~(0.05) \\
                                          & \gls{ast-lab}                &  &    0.530~(0.10) &    0.477~(0.12) & \B 0.871~(0.09)     &    0.514~(0.09) &    0.374~(0.09) \\
                                          & \gls{ast-seq}                &  & \B 0.597~(0.06) &    0.475~(0.10) &    0.870~(0.09)     & \B 0.591~(0.10) & \B 0.409~(0.07) \\ \midrule

\multirow{7}{*}{\acs{ldc} ($K=10$)}       & \acs{lr}                     &  &    0.048~(0.01) &    0.628~(0.16) &    0.722~(0.10)     &    0.010~(0.06) &    0.043~(0.01) \\
                                          & \acs{svm}                    &  &    0.161~(0.07) &    0.018~(0.01) &    0.711~(0.08)     &    0.082~(0.03) &    0.017~(0.01) \\
                                          & \acs{xgb}                    &  &    0.520~(0.14) &    0.068~(0.03) &    0.776~(0.07)     &    0.217~(0.09) &    0.065~(0.03) \\
                                          & \acs{mlp}                    &  &    0.461~(0.20) &    0.114~(0.05) &    0.825~(0.06)     &    0.254~(0.10) &    0.105~(0.05) \\
                                          & AlexNet                      &  &    0.579~(0.19) &    0.301~(0.10) &    0.888~(0.04)     &    0.390~(0.11) &    0.248~(0.08) \\
                                          & ResNet                       &  &    0.523~(0.19) &    0.276~(0.10) &    0.883~(0.04)     &    0.358~(0.11) &    0.221~(0.07) \\
                                          & \gls{ast-lab}                &  &    0.543~(0.13) &    0.380~(0.10) &    0.918~(0.03)     &    0.434~(0.11) &    0.295~(0.07) \\
                                          & \gls{ast-seq}                &  & \B 0.586~(0.13) & \B 0.411~(0.12) & \B 0.952~(0.03)     & \B 0.491~(0.12) & \B 0.328~(0.10) \\ \bottomrule
\end{tabular}

\end{figure}

As discussed in \cref{sec:call-det-end-class}, elephant call detection can also be viewed as a multi-label task, where not just the presence of a call is detected, but in the same step also the type of call.
\Cref{fig:proc:frame:call} compares the purity-coverage curves for a \ac{lr} and \gls{ast-seq} models in this per-frame multi-label scenario.
We observe that while the \ac{lr} performs well for \texttt{rumble} detection, the \gls{ast-seq} model offers vastly superior performance for the \texttt{roar} and \texttt{trumpet} classes.
Furthermore, we note that none of the classifiers could correctly detect more than \qty{50}{\percent} of the \texttt{cry} segments, reflected by the coverage score.
This may be due to the short duration and high energy present in the call type.
As a result of the acoustic properties of the call, there is no harmonic or formant structure to the call type.
This may also be because of the small number of samples available (\cref{fig:dataset:call_nseg}), where there are roughly only two exemplars per cross-validation fold.

The sudden jump in purity is the result of the classifier correctly identifying a large portion of frames after a small change in the classification threshold.
We found that for these exemplars, the acoustic features showed a high degree of correlation~(cosine similarity) between frames.
As a result, we expect these features to be close in the respective classifier feature space.

From both \cref{tab:results:detection:call} we observe that while the shallow classifiers achieved commendable performance in the task of binary detection, these models perform far worse than the deeper architectures for the multi-label tasks.
The \gls{ast}[-based] models, however, remain the top performing candidates.

\subsection{Call classification}
Next we assess the classifier's performance on a segment-level elephant call classification task in a multi-label scenario.
As stated before, it is assumed this task is downstream from a oracle endpointing model, and thus the segment location in time is computed from the annotation labels.
Thus, this task assess how well a model is able to classify an entire segment of audio.

\begin{figure}
  \begin{subfigure}{1.0\linewidth}
    \centering
    \caption{
      \Acs{elev} dataset
    }
    \label{fig:roc:seq:call:elev}
    \includegraphics[scale=0.85]{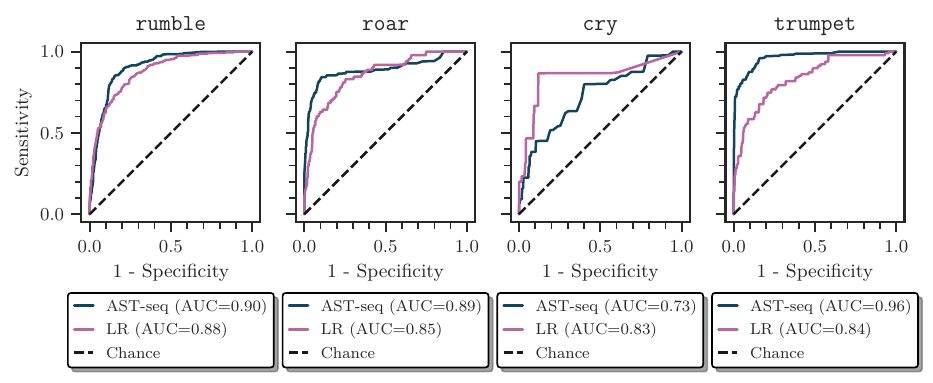}
  \end{subfigure}
  \begin{subfigure}{1.0\linewidth}
    \centering
    \caption{
      \Acs{ldc} dataset
    }
    \label{fig:roc:seq:call:ldc}
    \includegraphics[width=\textwidth]{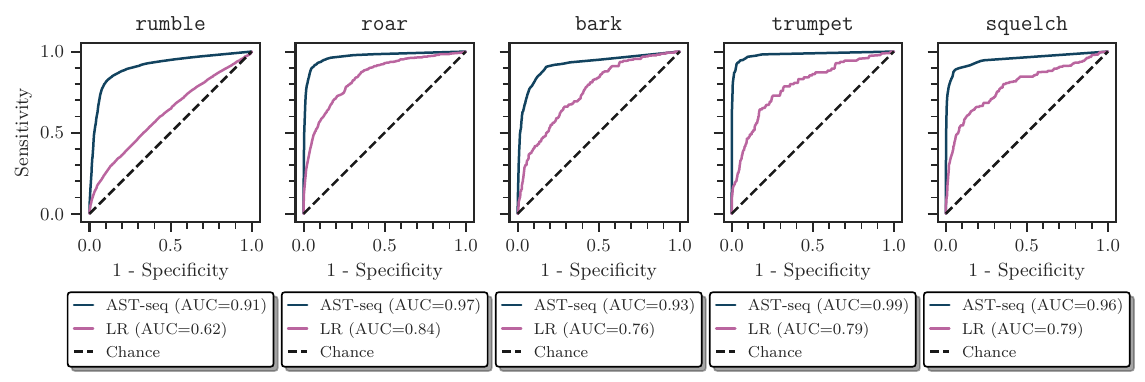}
  \end{subfigure}
  \caption{
    \Acf{roc} curves, for the task of segment-level elephant multi-label call classification, both \acf{elev} and \acf{ldc} datasets.
    The baseline model~(\ac{lr}) and the top performing model~(\gls{ast-seq}) are shown.
    The prevalence, which is the performance achieved when making random decisions according to the prior.
  }
  \label{fig:roc:seq:call}

\centering
\captionof{table}{
  Segment-level elephant multi-label call classification results for the \acf{elev} and the \acf{ldc} datasets for all considered models.
  The reported metrics are averages over the $K$ outer folds.
  The associated standard deviation is given in parentheses.
}
\label{tab:results:classification:call}

\sisetup{
  detect-weight,
  mode=text,
  table-alignment-mode = format,
  table-format = 1.3,
  table-number-alignment = center,
  separate-uncertainty = false,
}

\renewrobustcmd{\bfseries}{\fontseries{b}\selectfont}
\renewrobustcmd{\boldmath}{}

\newrobustcmd{\B}{\bfseries}

\begin{tabular}{lllccccc}
\toprule
Dataset                                   & Model                        & &    {Precision}  &    {Recall/Sens.} &    {Specificity}  &     {\Acs{auc roc}} \\ \midrule
\multirow{7}{*}{\acs{elev} ($K=5$)}       & \acs{lr}                     & &    0.325~(0.13) & \B 0.525~(0.27)   &    0.908~(0.06)   &     0.849~(0.12)    \\
                                          & \acs{svm}                    & &    0.453~(0.17) &    0.257~(0.06)   &    0.947~(0.03)   &     0.876~(0.09)    \\
                                          & \acs{xgb}                    & &    0.416~(0.15) &    0.245~(0.05)   &    0.951~(0.02)   &     0.866~(0.08)    \\
                                          & \acs{mlp}                    & &    0.436~(0.19) &    0.304~(0.12)   &    0.942~(0.02)   &     0.880~(0.08)    \\
                                          & AlexNet                      & &    0.492~(0.23) &    0.373~(0.13)   &    0.944~(0.02)   &  \B 0.920~(0.05)    \\
                                          & ResNet                       & &    0.406~(0.25) &    0.269~(0.07)   &    0.932~(0.05)   &     0.881~(0.04)    \\
                                          & \gls{ast-lab}                & &    0.529~(0.10) &    0.466~(0.12)   &    0.954~(0.03)   &     0.876~(0.12)    \\
                                          & \gls{ast-seq}                & & \B 0.533~(0.08) &    0.458~(0.12)   & \B 0.957~(0.01)   &     0.871~(0.09)    \\ \midrule

\multirow{7}{*}{\acs{ldc} ($K=10$)}       & \acs{lr}                     & &    0.119~(0.05) & \B 0.640~(0.20)   &    0.744~(0.11)   &    0.759~(0.12)     \\
                                          & \acs{svm}                    & &    0.200~(0.18) &    0.001~(0.01)   & \B 0.999~(0.01)   &    0.758~(0.10)     \\
                                          & \acs{xgb}                    & &    0.409~(0.17) &    0.039~(0.03)   & \B 0.999~(0.01)   &    0.775~(0.07)     \\
                                          & \acs{mlp}                    & &    0.470~(0.25) &    0.095~(0.06)   &    0.996~(0.00)   &    0.867~(0.07)     \\
                                          & AlexNet                      & &    0.638~(0.17) &    0.320~(0.15)   &    0.993~(0.01)   &    0.929~(0.05)     \\
                                          & ResNet                       & &    0.653~(0.26) &    0.302~(0.15)   &    0.991~(0.01)   &    0.916~(0.05)     \\
                                          & \gls{ast-lab}                & &    0.690~(0.28) &    0.390~(0.13)   &    0.988~(0.01)   &    0.954~(0.03)     \\
                                          & \gls{ast-seq}                & & \B 0.730~(0.20) &    0.435~(0.13)   &    0.987~(0.01)   & \B 0.957~(0.03)     \\ \bottomrule
\end{tabular}
\end{figure}

We observe that the models that perform well at per-frame detection~(including negative class) are also the models that perform well at segment-level classification~(excluding negative class) tasks, seen by comparing the results in \cref{tab:results:detection:call,tab:results:classification:call}.

Furthermore, we observe the same trend in segment-level classification of low-resource classes, such as \texttt{cry}, that we saw for per-frame detection.
\Cref{fig:roc:seq:call}, shows how the worst performing class remains the \texttt{cry}-class.
However, in the case of the segment-level classification, the \ac{ast} and other deep architectures all overfit on the \texttt{cry}-class, while the shallow classifiers showed superior performance.
In contrast, the opposite is true when evaluating the models in a multi-label per-frame detection scenario.
Such overfitting could in future be addressed by employing low-resource techniques such as few-shot learning, where a nearest neighbour approach is used in conjunction on the feature representation obtained from pre-trained neural network.

\subsection{Subcall classification}
As discussed in \cref{sec:subcalls}, elephant produce a set of nuanced calls that have been shown to strongly correlate with animal behaviour.
The ability to automatically identify such subcalls would greatly benefit animal monitoring and provide near-realtime indications on herd health and status.

As before, we employ several shallow and deep classification models in order to perform segment-level classification of these subcalls.
In order to maintain the nested cross-validation experimental setup, only a subset of the subcalls can be evaluated~(\cref{fig:dataset:subcall_nseg}).
Furthermore, only the \ac{elev} dataset can be used since only it contains the required subcall annotations.
Hence, the \ac{ldc} dataset is not used in the subcall experiment.

\begin{figure}

  \centering
  \includegraphics[scale=0.85]{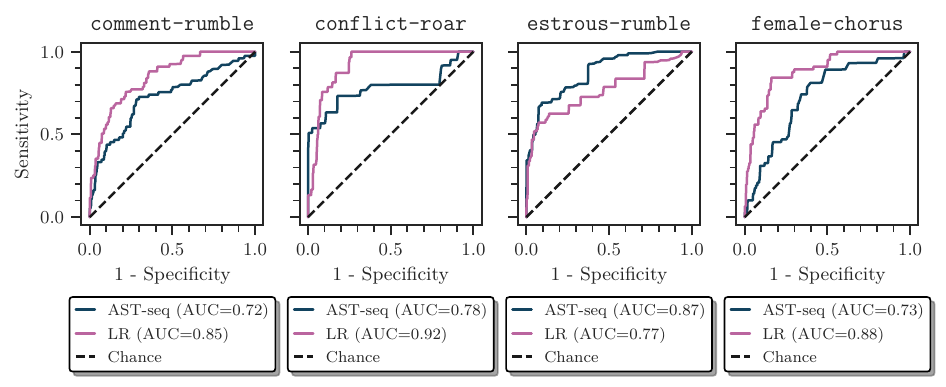}
  \includegraphics[scale=0.85]{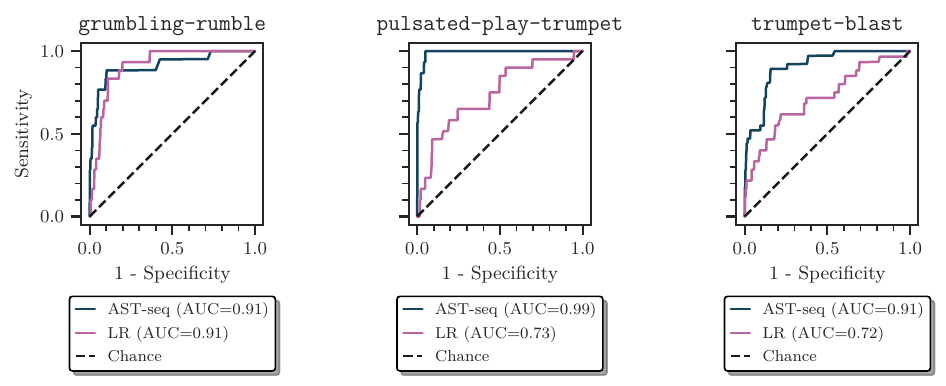}
  \caption{
    \Acf{roc} curves, for the task of segment-level elephant multi-label subcall classification, for \acf{elev} dataset.
    The baseline model~(\ac{lr}) and the top performing model~(\gls{ast-seq}) are shown.
    The prevalence, which is the performance achieved when making random decisions according to the prior.
  }
  \label{fig:roc:seq:subcall}

\centering
\captionof{table}{
  Segment-level elephant multi-label subcall classification results for the \acf{elev} dataset for all considered models.
  The reported metrics are averages over the $K$ outer folds.
  The associated standard deviation is given in parentheses.
}
\label{tab:results:classification:subcall}

\sisetup{
  detect-weight,
  mode=text,
  table-alignment-mode = format,
  table-format = 1.3,
  table-number-alignment = center,
  separate-uncertainty = false,
}

\renewrobustcmd{\bfseries}{\fontseries{b}\selectfont}
\renewrobustcmd{\boldmath}{}

\newrobustcmd{\B}{\bfseries}

\begin{tabular}{lllccccc}
\toprule
Dataset                                   & Model                        & &    {Precision}  &    {Recall/Sens.} &    {Specificity}  &    {\Acs{auc roc}} \\ \midrule
\multirow{7}{*}{\acs{elev} ($K=5$)}       & \acs{lr}                     & &    0.141~(0.08) & \B 0.680~(0.03)   &    0.824~(0.12)   &    0.826~(0.10)    \\
                                          & \acs{svm}                    & &    0.029~(0.06) &    0.007~(0.02)   &    0.992~(0.02)   &    0.826~(0.14)    \\
                                          & \acs{xgb}                    & &    0.107~(0.15) &    0.037~(0.07)   & \B 0.999~(0.00)   &    0.864~(0.12)    \\
                                          & \acs{mlp}                    & &    0.131~(0.20) &    0.056~(0.08)   & \B 0.999~(0.00)   &    0.875~(0.12)    \\
                                          & AlexNet                      & &    0.395~(0.36) &    0.159~(0.14)   &    0.998~(0.00)   &    0.888~(0.09)    \\
                                          & ResNet                       & &    0.246~(0.30) &    0.116~(0.14)   &    0.998~(0.00)   &    0.875~(0.09)    \\
                                          & \gls{ast-lab}                & & \B 0.613~(0.31) &    0.325~(0.20)   &    0.996~(0.01)   & \B 0.979~(0.10)    \\
                                          & \gls{ast-seq}                & &    0.374~(0.27) &    0.251~(0.23)   &    0.993~(0.01)   &    0.845~(0.12)    \\ \bottomrule
\end{tabular}
\end{figure}

\Cref{fig:roc:seq:subcall} shows is the \ac{roc} curves associated with the results presented in \cref{tab:results:classification:subcall}.
As before, we observe that the shallow models, while strong contenders in the binary per-frame detection task, do not perform well in these multi-label tasks~(including subcall classification).

Due to the small number of samples available to train the deeper models, overfitting occurs.
However, in contrast to segment-level call classification, where the shallow models performed better on these low resources classes~(such as \texttt{cry}) this is not the case for subcall classification.
We postulate that subcalls may in fact be a more difficult classification problem, thus while the deeper architectures may overfit, when incorporating a pre-trained network this risk is reduced.
Thus, we ultimately use the deep neural networks as feature extractors~(encoder structure) and rely on the final linear layer to perform the actual classification.

As before, we see that the transformer models achieve best performance followed by the \ac{cnn}[-based] models, and then \ac{mlp}.

\section{Discussion}

\subsection{Feature space analysis}
The results of our multi-label per-frame detection experiments suggest that using speech-based features as input might be limiting the performance of certain models.
Although the effectiveness and applicability of these features beyond speech research is contested, however these features are considered essential when utilising pre-trained models.
To evaluate this hypothesis, we compare the high-dimensional \ac{mfcc}[-] and mel-spectral features using a \ac{umap} as a dimensionality reduction.
We compare these input features along with the pre-trained \ac{ast} representations.

We obtain the \ac{ast} representations by computing the output of the last layer of the transformer~(before the \ac{mlp} classification head) for a given input context.
As the model produces a sequence of outputs both for the spectral and temporal input dimensions, we average the outputs associated with the spectral input dimension to obtain a single 768-dimensional vector every \qty{160}{\milli\second} in time.

Each of the respective high-dimensional features and representations are reduced to a two-dimensional projection using a \ac{umap}~\autocite{mcinnes2020umap}.
The \ac{umap} is calibrated using the training labels, and applied to the development set.

\begin{figure}
  \centering
  \includegraphics{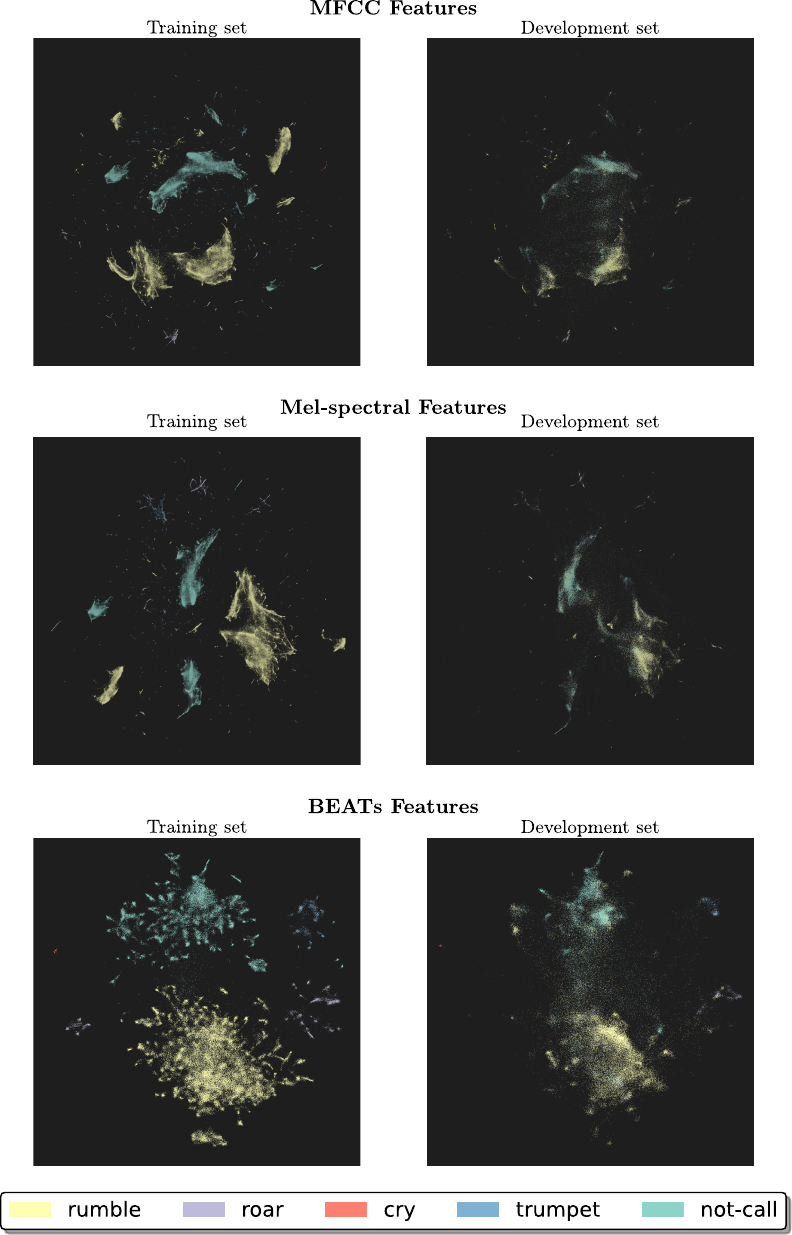}
  \caption{
    Supervised \acs{umap} of \acs{mfcc}, mel-spectral and \acs{beats} features for a particular inner cross-validation turn.
  }
  \label{fig:umap}
\end{figure}

\Cref{fig:umap} shows the two-dimensional \ac{umap} for a single inner cross-validation fold of the \ac{elev} dataset.
\Ac{mfcc}, mel-spectral features, and \ac{ast} representations were extracted from both training and development fold.
It is evident that, while discernible clusters form in the training projections for all three feature types, the spectral features~(mel and \ac{mfcc}) do not generalise well to the development fold.
Despite this, a clear separation between the two classes \texttt{no-call} and \texttt{rumble} is observed, for all features.
This could explain the high performance of shallow models in binary per-frame detection but poor performance in multi-label per-frame detection.
The \ac{ast} features do however generalise well to the validation fold and maintain similar cluster shape.

Upon examination of misclassifications made by the shallow models, it was found that the classifier struggled with highly polyphonic segments.
These polyphonic instances typically include noise, such as engine sounds, wind rustling foliage; or other dominant sound sources such as birds, insects, or combinations thereof.
Identifying the precise call boundary for such instances proved challenging even for human annotators.
We believe that the low-energy tails of such vocalisations were only noticed by the annotators due to the high-energy start of the vocalisation.
Additional misclassifications were attributed to recording artefacts, such as microphone handling sounds and unanticipated starting or ending instances within recordings.
It is noteworthy that some misclassifications might even be considered legitimate, as certain ground truth labels marginally fall short of the threshold required for a positive sample classification.

Deep-classification models with larger temporal contexts faced similar challenges, although resulting in fewer misclassifications in the presence of polyphonic source environments.
However, the \ac{cnn} models still misclassify low-energy vocalisations with extended tails.
The \gls{ast}[-based] models were the only classifiers that could accurately detect low-energy vocalisation in a highly polyphonic settings, likely due to the global attention mechanism present in its transformer-encoder architecture.

\subsection{Attention map analysis}
We now will consider the attention map used by the \gls{ast-seq} model for a particular exemplar drawn from a development fold of the \ac{elev} dataset.
The transformer attention mechanism, as briefly described, allows to model to selectively focus on particular features in the input, based on other input features.
This characteristic of the attention mechanism enables contextualised classifications based on past and future features in the extended input context.
We postulate that this is the reason for the models improved ability to detect low-energy vocalisations amidst other acoustic sources.

\begin{figure}
  \centering
  \includegraphics{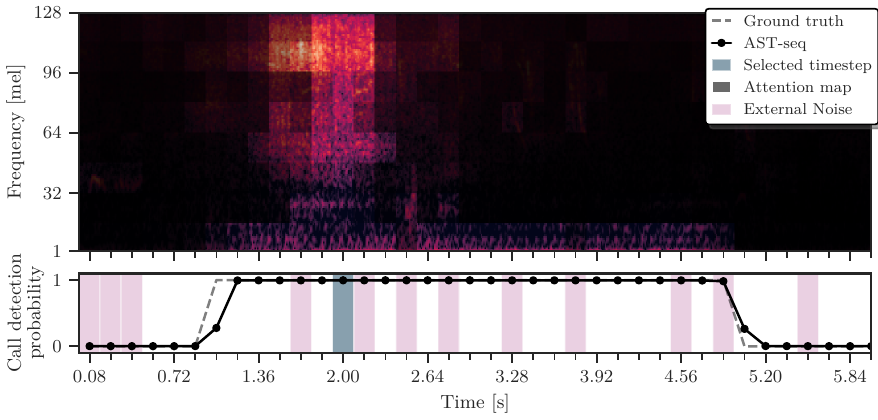}
  \caption{
    Mel-spectrogram input to \gls{ast-seq} model with binary call detection probability output shown for timestep \qty{2}{\second}.
    Attention map overlay shown, darker regions of the attention map shows features rejected by the attention mechanism, and vice versa.
    In addition external noise sources~(e.g. bird calls) are also indicated.
    A detection probability is produced per temporal patch every \qty{160}{\milli\second}.
    The exemplar is drawn from the development set of the given cross-validation fold.
  }
  \label{fig:attention-map}
\end{figure}

\Cref{fig:attention-map} shows the attention map for one isolated model output~(token), in a binary call detection scenario.
Note that the only model architecture that correctly identified this example segment was \gls{ast-seq}.
From the attention map we can see that the model uses features from both the past and the future to produce the chosen classification output.
The majority of the input features are masked by the attention map, but not for the region corresponding to the low-frequency elephant rumble.
We also see that features from other external noise sources are correctly rejected.
The focused attention allows the model to better reject acoustic sources that may have an adversarial effect on the model performance.

From \cref{fig:attention-map} we see that the \gls{ast-seq} model is also attending to acoustic sources other than elephants, e.g.\ bird calls~(indicated as \emph{external noise} in the figure).
This could mean that the model is using these external audio sources as a proxy for elephant vocalisation, or that the model is using these calls to reject other external sources.

With the limited data available in the \ac{elev} dataset is it not yet possible to discern the mechanism at play.

\subsection{Classifier performance}
Here, we broadly discuss the overall tends observed by classifier performance across both datasets and the various tasks reported in \cref{sec:results}.

First, we compare \ac{lr} and \ac{svm}.
These models have similar computation complexity and both rely on a hyperplane decision boundary.
Due to the non-linear kernel function of the \ac{svm} model, the decision boundary in no longer linear as is the case with \ac{lr}.
As a results, we observe that the \ac{svm} model outperforms the \ac{lr} models in the majority of cases.
However, this additional kernel function comes at an increased computational and memory complexity.
Furthermore, if an output probability is required, the \ac{svm} classifier requires an additional classifier calibration step.
Thus, as a lightweight model for early preprocessing of binary per-frame detection~(for later downstream tasks), \ac{lr} is a model that has computational cost to performance balance.

\Ac{xgb} had similar performance to the \ac{mlp} models.
However, the hyperparameter search was far more computationally expensive to perform for \ac{mlp}.
This is in part due to the large hyperparameter space that the \ac{mlp} models occupy and how increases in the depth or width of the model increase its the computational complexity.
In contrast, the \ac{xgb} model hyperparameters had less variance over the search, indicating that the model was more robust to na\"ive model parameter selection.
Furthermore, many of the \ac{xgb} model parameters result in a reduction in the model capcity, by reducing for instance the number of leaf nodes or trees.
As such, searching the hyperparameter space generally leads to computational improvements.

When considering the \ac{cnn}[-based] models, we observe that overall AlexNet either exceeded or matched the ResNet performance.
This is counter intuitive, as the residual connections in ResNets allow for deeper models to be trained.
However, due to the small size of the datasets used in this work, we do not observe these benefits to performance in our experimental results.

Finally, we compare the two best performing models: \gls{ast-lab} and \gls{ast-seq}.
The only difference between these transformer-based models is in their output layer.
However, we observe in some cases that there is a noticeable performance difference between these two models.
For most endpointing metrics, the \gls{ast-seq} model is superior.
We postulate that this is due to the sequence-to-sequence training regime, where the explicit call segment boundaries are presented to the model during training, while for the \gls{ast-lab} model this has to be implicitly learned as it is only given a single output target.
However, this does also make detections near the start and the end of the input context more difficult for the \gls{ast-seq} model.
While the \gls{ast-lab} model always produces a model classification of the centre of the input context, the \gls{ast-seq} model is tasked with producing classification outputs near the start and the end of the same input context, with less local contextual information.

Hence \gls{ast-seq} is better at better at call endpointing, while \gls{ast-lab} is better at detection and classification.

\section{Summary and Conclusion}
We have considered the automatic detection, endpointing and classification of elephant vocalisations in audio recordings.
A combination of both shallow and deep architectures were evaluated against two audio datasets containing elephant vocalisations.
These datasets consist of multi-label classifications targets.
By employing classifiers on a per-frame basis, we are able to compute the multi-label call probability every \qty{100}{\milli\second}.
From these per-frame detections we are able to localise the call in time and endpoint the start and the end of the call in the recording.
Furthermore, we evaluate the models ability to identify the an unknown elephant vocalisation from a endpointed audio segment.

We have evaluated \ac{lr}, \ac{svm} and \ac{xgb} as our chosen shallow model architectures, and \ac{mlp}, \ac{cnn}[-based] AlexNet and ResNet architectures and finally the transformer-based \ac{ast} models as the chosen deep model architectures.
These models have been evaluated on a large range of hyperparameters in an attempt to find the best selection of model parameters for the given datasets.

We have also investigated the use of a dimensionality reduction pre-processing step and \ac{mfcc} and mel-spectral feature configurations.
For shallow models, an input context length of \qtyrange{500}{1000}{\milli\second} performed best, followed by a \ac{pca} dimensionality reduction step.
For the \ac{cnn}[-based] models, an input length of \qty{2.5}{\second} lead to the best performance.

For the deep architectures, transfer learning using out-of-domain pre-trained networks was found to lower the overall training time but lead to only minor or no performance improvements compared to starting from random initialisation.
In contrast, using in-domain pre-trained models, lead to both a reduction in training time and improved model performance.
The application of task transfer by finetuning a model pre-trained to produce a single classification output to produce a sequence of classification outputs drastically improves the computational efficiency of the model, and also led to the best performing model~(\gls{ast-seq}).

The \ac{ast}[-based] models performed the best in all three tasks: call detection (both binary and multi-label), call endpointing and call classification.
However we also find that \ac{lr} is a viable lightweight model for binary call detection.
In both multi-label call detection and classification tasks, it was found that the shallow classification models lead to a deterioration in performance, compared to binary call detection.
We speculate this is because the \ac{mfcc} input feature space used by the shallow models not representative of the call characteristics.

In conclusion, this study has demonstrated the significant benefits of pre-training and ensuant success of sophisticated models, in the context of small bioacoustic datasets.
The \gls{ast-seq} model emerged as a top performer in terms of both computational efficiency, due to the sequence-to-sequence structure, and detection as well as classification performance, on both framewise and sequence tasks.
Additionally, we found that, while speech features perform well in detection tasks, they are not as effective for classification purposes.
Finally, transformer models offer a promising avenue for sub-call classification, demonstrating their successful application in early automated animal behavioural classification.

\section{Acknowledgements}
We gratefully acknowledge financial support by Telkom South Africa for the research presented in this paper.
We would also like to thank the Stellenbosch Rhasatsha High Performance Computing facility and team for access to their facilities and technical support.
Their expertise and resources were invaluable to the successful completion of this project.
Finally, We would like to acknowledge NVIDIA for their generous donation of \acp{gpu} which provided additional computational resources which were key to the training and evaluation of the neural network classifiers.
\clearpage

\printbibliography[sorting=nyt]
\end{document}